\documentclass[aps,prd,showpacs,amssymb,floatfix,twocolumn,prd,pdftex]{revtex4}
\usepackage{graphicx}
\usepackage[sort&compress]{natbib}
\usepackage{mathrsfs}
\begin{document}
\newcommand{\eg}{{\it e.g.}}
\newcommand{\etal}{{\it et. al.}}
\newcommand{\ie}{{\it i.e.}}
\newcommand{\be}{\begin{equation}}
\newcommand{\dd}{\displaystyle}
\newcommand{\ee}{\end{equation}}
\newcommand{\bea}{\begin{eqnarray}}
\newcommand{\eea}{\end{eqnarray}}
\newcommand{\bef}{\begin{figure}}
\newcommand{\eef}{\end{figure}}
\newcommand{\bce}{\begin{center}}
\newcommand{\ece}{\end{center}}
\def\lsim{\mathrel{\rlap{\lower4pt\hbox{\hskip1pt$\sim$}}
    \raise1pt\hbox{$<$}}}         
\def\gsim{\mathrel{\rlap{\lower4pt\hbox{\hskip1pt$\sim$}}
    \raise1pt\hbox{$>$}}}         


\title{Quark deconfinement transition in neutron stars with the field correlator method}
  \author{Domenico Logoteta\,$^1$ and  Ignazio Bombaci\,$^2$ }
  \affiliation{$^1$Centro de F\'{\i}sica Computacional, Department of Physics,
   University of Coimbra, 3004-516 Coimbra, Portugal}
  \affiliation{$^2$Dipartimento di Fisica ``Enrico Fermi'', Universit\'a di Pisa,
 and INFN Sezione di Pisa,  Largo Bruno Pontecorvo 3, I-56127 Pisa, Italy }

\received{21 May 2013;  accepted 19 August 2013}

\begin{abstract} 
A phase of strong interacting matter with deconfined quarks is expected in the core of massive neutron stars. 
In this article, we perform a study of the hadron-quark phase transition in cold ($T = 0$) 
neutron star matter and we calculate various structural properties of hybrid stars.   
For the quark phase, we make use of an equation of state (EOS) derived with 
the field correlator method (FCM) recently extended to the case of nonzero baryon density. 
For the hadronic phase, we consider both pure nucleonic and hyperonic matter, and we derive 
the corresponding EOS within a relativistic mean field approach.    
We make use of measured neutron star masses, and particularly the mass $M = 1.97 \pm 0.04 \, M_\odot$ 
of PSR~J1614-2230 to constrain the values of the gluon condensate $G_2$, which is one of the 
EOS parameters within the FCM.  
We find that the values of $G_2$ extracted from the mass measurement of PSR~J1614-2230 
are consistent with the values of the same quantity derived within the FCM  
from recent lattice QCD calculations of the deconfinement transition temperature at 
zero baryon chemical potential. 
The FCM thus provides a powerful tool to link numerical calculations of QCD on 
a space-time lattice with measured neutron star masses.   
\end{abstract}

\maketitle

------------------------------------------------------------

\section{Introduction}
\label{intr} 
Neutron stars contain in their interiors one of the densest form of matter in the Universe. 
In fact, neutron star structure calculations \cite{lp01,peng08,li08,vida11}  
based on a large variety of modern equations of state of dense hadronic matter predict a maximum 
stellar central density (the one for the maximum mass star configuration) in the range of  4 -- 8 times 
the saturation density ($\sim 2.8 \times 10^{14}$~g/cm$^{3}$) of nuclear matter.  
Thus these stars can be viewed as natural laboratories to explore the low-temperature $T$ and 
high baryon chemical potential region of the phase diagram of quantum chromodynamics (QCD) 
\cite{alf08,web05,prak97,bom07}.    
In this regime nonperturbative aspects of QCD are expected to play a crucial role, and a transition 
to a phase with deconfined quarks and gluons is expected to occur and to influence a number of 
interesting astrophysical phenomena \cite{pg10,sot11,r1,r3,r6,r7,weis11,nish12}. 

Recent high-precision numerical calculations of QCD on a space-time lattice at zero baryon chemical 
potential $\mu_b$ (zero baryon density) have shown that at high temperature 
and for physical values of the quark masses, the transition to quark gluon plasma is a 
crossover \cite{bern05,cheng06,aoki06} rather than a real phase transition.  

Unfortunately, present lattice QCD calculations at finite baryon chemical potential are 
plagued with the notorious ``sign problem'', which makes them unrealizable by all 
presently known lattice methods (see e.g. \cite{mp_lomb08} and references therein).     
Thus, to explore the QCD phase diagram at low-temperature T and high $\mu_b$, it is 
necessary to invoke some approximations in QCD or to apply a QCD effective model.   

Along these lines, for example, a model of the equation of state (EOS) 
of quark matter \cite{mit-eos} inspired by the MIT bag model of hadrons \cite{mit} 
has been intensively used by many authors to calculate the structure of 
strange stars \cite{witt84,afo86,hzs86,XDLi99,Xu99}, or the structure of the 
so-called hybrid stars, i.e. neutron stars with a quark matter core.   
In this model quark matter is treated as a free relativistic Fermi gas of {\it u}, {\it d} and  {\it s} 
quarks, which reside in a region characterized by a constant energy density $B$, with at most   
perturbative corrections up to the second order in the QCD structure constant $\alpha_s$ 
\cite{FMc77,Bal78,Kur10}. The parameter $B$ takes into account, in a crude phenomenological 
manner, nonperturbative aspects of QCD and it is related to the bag constant which 
in the MIT bag model \cite{mit} gives the confinement of quarks within hadrons.   
The bag model EOS is expected to be reasonable at asymptotically large density,  but 
it crumbles in the density region where quarks clusterize to form hadrons, i.e.  
in the region where the deconfinement phase transition takes place. 

Another very used quark  model is the Nambu--Jona-Lasinio (NJL) model  \cite{njl} 
(for a thorough review see \cite{bub05}). This model has proved to be very successful in the  
description of the spontaneous breakdown of chiral symmetry exhibited by true QCD vacuum.  
It explains very well the spectrum of the low-lying mesons as well as many other low-energy 
phenomena of strong interaction \cite{sl1,sl2,sl3}.    
The NJL is not a confining model and it is based on an effective chiral Lagrangian that 
captures some of symmetries of QCD. 
In the NJL approach quarks interact each other through a nonrenormalizable pointlike  
Fermi interaction. The entire system is gluon free so it cannot be used in the limit of 
low density and high temperature. 

In the case of the MIT bag model a large window of the parameters allows for the 
existence of stable strange stars or hybrid stars.   
In the case of the NJL model the existence of stable hybrid stars is very unlikely \cite{benh1}  
or is possible only for a restricted range of the model parameters 
\cite{sche99,bub,yang08, log12}.   
   
As we have commented, the MIT bag model and the NJL model (as other QCD effective models)  
cannot make predictions in the high-$T$ and zero baryon chemical potential region, and thus 
cannot be tested using present lattice QCD calculations. 

Recently the deconfinement phase transition has been described using an EOS of quark 
gluon plasma derived within the field correlator method (FCM) \cite{st,digiac02}  
extended to finite baryon chemical potential \cite{ST07,sim1,sim3,sim5}. 
The field correlator method  is a nonperturbative approach to QCD 
which includes from first principles the dynamics of confinement 
in terms of color electric and color magnetic correlators. 
The model is parametrized in terms of the gluon condensate $G_2$ 
and the large-distance static quark-antiquark ($Q\bar{Q}$) potential $V_1$.  
These two quantities control the EOS of the deconfined phase at fixed quark masses and temperature. 
The main constructive characteristic of the FCM is the possibility to describe the whole 
QCD phase diagram  as it can span from high temperature and low baryon chemical potential,  
to low $T$ and high $\mu_b$ limit. 

A very interesting feature of the FCM is that the value of the gluon condensate can 
be obtained  from lattice QCD calculations \cite{bors10,baz12} of the deconfinement transition 
temperature $T_c$, at zero baryon chemical potential.  
Thus we have an efficacious tool to directly link lattice simulations and neutron star physics.    

To explore this link is one of the main purposes of the present work. 
In particular, we will investigate the possibility for the occurrence of the quark 
deconfinement transition in neutron stars and the possibility to have stable hybrid star 
configurations using the field correlator method for the quark phase EOS and a 
relativistic mean field model \cite{wale,bog77} for the EOS of the hadronic phase.      
A similar study has been performed in Ref.~\cite{Baldo}, where a microscopic EOS derived with 
the Brueckner-Hartree-Fock approximation \cite{bbb97,bal90,zhLi06} has been employed to describe 
the hadronic phase, whereas, for the quark phase, the FCM has been used as in the present work.
The properties of absolutely stable \cite{witt84} strange quark matter and strange stars 
have been recently investigated within the FCM by the author of Ref.~\cite{Pereira}.   

This work is organized as follows: In Sec.~\ref{eosqm} we briefly review the FCM at finite 
temperature and density; in Secs.~\ref{eoshm} and \ref{phtr} we discuss respectively 
the hadronic EOS and the formalism of phase transition to quark matter in 
$\beta$-stable hadronic matter; our main results are presented in Sec.~\ref{NSstr}; 
the link between lattice QCD calculations and measured neutron star masses is discussed 
in Sec.~\ref{Tc};  finally the conclusions of this work are outlined in Sec.~\ref{concl}.

\section{EOS of the Quark Phase}
\label{eosqm}

The quark matter equation of state we used in the present work is based on the FCM \cite{st} 
(see \cite{digiac02}for a detailed review). 
Recently this method has been extended to the case of nonzero baryon 
density \cite{ST07,sim1,sim3,sim5} making possible its application to neutron star matter. \\
The principal advantage of the FCM is a natural explanation and treatment of the dynamics of 
confinement in terms of color electric  $D^{E}(x)$, $D_{1}^{E}(x)$ and color magnetic  
$D^{H}(x)$, $D_{1}^{H}(x)$ Gaussian correlators. 

The correlators $D^{E}(x)$, $D_{1}^{E}(x)$ and  $D^{H}(x)$, $D_{1}^{H}(x)$  
are related to the nontrivial two-point correlation function for the color electric and 
color magnetic fields \cite{digiac02}.  \\  
$D^{E}$ contributes to the standard string tension $\sigma^{E}$ through \cite{ST07}  
\be
\label{f1}
\sigma^{E}=\frac{1}{2}\int D^{E}(x) \ d^{2}x .
\ee
The string tension $\sigma^{E}$ vanishes as $D^{E}$ goes to zero at $T \geq T_c$ 
and this leads to deconfinement. 
The correlators have been calculated on the lattice \cite{pisa1,pisa2,pisa3} 
and also analytically \cite{sim98}. 

In the lowest nonperturbative approximation one can hold only single quark and gluon interactions 
with the vacuum. This is the so-called called single line approximation \cite{ST07}.   
Using this approximation the quark pressure $P_q$ for a single flavor, reads  
\cite{ST07,sim3,sim5} 

\be\label{pquark}
P_q/T^4 = \frac{1}{\pi^2}[\phi_\nu (\frac{\mu_q - V_1/2}{T}) +
\phi_\nu (-\frac{\mu_q + V_1/2} {T})]
\ee
where 
\be
\label{eq:phi}
\phi_\nu (a) = \int_0^\infty du \frac{u^4}{\sqrt{u^2+\nu^2}} 
               \frac{1}{\exp{[\sqrt{u^2 + \nu^2} - a]} + 1}\, , 
\ee
$\nu=m_q/T$ and  $V_1$ is the large-distance static $Q\bar Q$ potential:
\be
\label{v1}
V_1 = \int_0^{1/T} d\tau(1-\tau T) \int_0^\infty d\chi \chi D_1^E(\sqrt{\chi^2 + \tau^2}) \, .
\ee
The nonperturbative contribution to  $D_1^E(x)$ is parametrized as \cite{digiac02} 
\be
\label{d1nonpt}
D_1^E(x)=D_1^E(0) \exp(-|x|/\lambda)
\ee
where $\lambda$ is the vacuum correlation length.  
Following Ref.~\cite{ST07}, we use the value  $\lambda = 0.34~\rm{fm}$ 
which has been determined in lattice QCD calculations \cite{pisa2}. 

In this formalism $V_{1}$ in Eq.(\ref{v1}) is independent on the chemical potential 
(and so on the density). This feature is partially supported by lattice simulations 
at small chemical potential \cite{ST07,latmuf}.  
In the present work, the value of $V_{1}$ at $T=0$ has been considered as a 
model parameter \cite{Baldo,Pereira}.    

The gluon contribution to the pressure is \cite{sim5} 
\be
\label{pglue}
P_g/T^4 = \frac{8}{3 \pi^2} \int_0^\infty  d\chi \chi^3
\frac{1}{\exp{(\chi + \frac{9 V_1}{8T} )} - 1}   \, .
\ee

In summary the total pressure of the quark phase is given by 
\be
\label{pq}
      P_{qg} = P_{g} + \sum_{u,d,s} P_{q} - \frac{9}{64} G_2 \, .
\ee
The last term in Eq.(\ref{pq}) represents the vacuum energy difference between 
the quark and hadronic phases, in the case of three-flavor ({\it u,\,d,\,s}) quark matter \cite{ST07},    
and $G_2$ is the gluon condensate. 
The latter quantity has been determined using QCD sum rules \cite{QCDsum-rules} to be in the range  
$G_2 = (0.012 \pm 0.006)~{\rm GeV}^4$. 
In the present work, following \cite{ST07,sim1,sim3,sim5}, we assume that the gluon condensate  
is independent on the baryon chemical potential, 
and we consider the value of $G_2$ as a model parameter.   

Notice that the quark pressure given in Eqs.~(\ref{pquark})--(\ref{d1nonpt}) is the one 
of a relativistic ideal Fermi gas, which in the case of $T = 0$ can be written as 
\bea
\label{P_FG0}
P_q &=& \frac{1}{4\pi^2} 
\Bigg\{ k_{F,q}^3 \sqrt{k_{F,q}^2 + m_q^2} 
   -\frac{3}{2}\;m_q^2 \Bigg[k_{F,q}\sqrt{k_{F,q}^2+m_q^2} \nonumber \\
&-& m_q^2\;\ln\Bigg(\frac{k_{F,q}+\sqrt{k_{F,q}^2+m_q^2}}{m_q}\Bigg)\Bigg]\Bigg\}\;,
\eea
where the Fermi momentum $k_{F,q}$ is related to the chemical potential $\mu_q$ 
of quarks with flavor $q$ by  
\be 
\label{mu_q}
    \mu_q = \sqrt{k_{F,q}^2 + m_q^2} + \frac{V_1}{2} \,. 
\ee
The energy density $\varepsilon_q$ at $T=0$ can be obtained using Eq.(\ref{P_FG0}) 
and the thermodynamical relation
\be
\varepsilon_q = - P_q + \mu_q\,n_q \,, 
\ee
where $n_q = \frac{1}{\pi^2}\, k_{F,q}^3$\, is the number density for quarks with flavor $q$.\\
One thus obtains the energy density of a relativistic ideal Fermi gas plus an extra term 
\be
  \varepsilon_q^{\prime} = \frac{V_1}{2}\, \frac{k_{F,q}^3}{\pi^2} 
\ee
which originates from the large-distance static $Q\bar Q$ potential $V_1 \equiv V_1(T=0)$. 

In our calculations we used the following values of the current-quark masses: 
$m_u = m_d = 5$~MeV and $m_s = 150$~MeV.   

In summary, the EOS for the quark gluon phase has two parameters: $G_2$ and $V_1 \equiv V_1(T=0)$. 
 
In the present work we have not considered the possibility of color superconductivity 
(see e.g. \cite{alf08} and references therein).  
As discussed in \cite{sim1,digia07} the low-temperature deconfinement transition produces a strong 
nonperturbative attraction in colorless channels that results in a dominance of the $Q\bar{Q}$ correlations 
over the diquark QQ ones. Thus within the FCM, QQ pairing and possible phases of color superconducting quark matter are hardly possible \cite{sim1,digia07}.  

\section{EOS of the Hadronic Phase}
\label{eoshm}

We adopt a nonlinear relativistic mean field model \cite{wale,bog77} for the EOS of hadronic 
matter and we make use of the  parametrization GM1 given by Glendenning and Moszkowski \cite{gm,gle}.   
The Lagrangian density including the baryonic octet, in terms of scalar $\sigma$, 
the vector-isoscalar $\omega_{\mu}$, and the vector-isovector $\vec{\rho_{\mu}}$ meson fields, reads 

\be
\label{l1}
\mathscr{L}=\mathscr{L}_{hadrons}+\mathscr{L}_{leptons} \ ,
\ee
where the hadronic contribution is
\be
\label{l2}
\mathscr{L}_{hadrons}=\mathscr{L}_{baryons}+\mathscr{L}_{mesons} \ ,
\ee
with
\be
\label{l3}
\mathscr{L}_{baryons}=\sum_{baryons} \bar{\Psi}_{B}[\gamma^{\mu} D_{\mu}-M_{B}^{*}]\Psi_{B} \ ,
\ee   
where
\be
D_{\mu}=i \partial_{\mu}-g_{\omega B} \omega_{\mu} - g_{\rho B} \vec{t}_{B}\cdot  \vec{\rho}_{\mu} \ ,
\ee
and $M_{B}^{*}=M_{B}-g_{\sigma B} \sigma$. The quantity $\vec{t}_{B}$ designates the isospin of baryon $B$. The mesonic contribution reads
\be
\label{l5}
\mathscr{L}_{mesons}=\mathscr{L}_{\sigma}+\mathscr{L}_{\rho}+\mathscr{L}_{\omega} \ ,
\ee
with
\be
\label{l6}
\mathscr{L}_{\sigma}=\frac{1}{2}(\partial_{\mu}\sigma\partial^{\mu}\sigma-m_{\sigma}^{2})+\frac{1}{3!}\kappa \sigma^{3}+\frac{1}{4!}\lambda\sigma^{4} \ ,
\ee

\be
\label{l7}
\mathscr{L}_{\omega}=-\frac{1}{4}\Omega_{\mu\nu}\Omega^{\mu\nu}+\frac{1}{2}m_{\omega}^{2}\omega_{\mu}\omega^{\mu} \ ,
\ee

\be
\label{l9}
\Omega_{\mu\nu}=\partial_{\mu}\omega_{\nu}-\partial_{\nu}\omega_{\mu} \ ,
\ee

\be
\label{l10}
\mathscr{L}_{\rho}=-\frac{1}{4}\vec{B}_{\mu\nu} \vec{B}^{\mu\nu}+\frac{1}{2}m_{\rho}^{2}\vec{\rho}_{\mu}\cdot  \vec{\rho}^{\mu} \ ,
\ee

\be
\label{l11}
\vec{B}_{\mu\nu}=\partial_{\mu}\vec{\rho}_{\nu}-\partial_{\nu}\vec{\rho}_{\mu}-g_{\rho}(\vec{\rho}_{\mu} \times \vec{\rho}_{\nu}) \ .
\ee
For the lepton contribution we take 
\be
\label{l12}
\mathscr{L}_{leptons}=\sum_{leptons} \bar{\Psi}_{l}[\gamma^{\mu} \partial_{\mu}-m_{l}^{*}]\Psi_{l} \ ,
\ee 
where the sum is over electrons and muons.\\ 

We have  used the parametrization of the nonlinear relativistic mean field model due 
to Glendenning and Moszkowski \cite{gm,gle}. The nucleon coupling constants are fitted 
to the bulk properties of nuclear matter. In particular, for the GM1 parametrization \cite{gm,gle}
the incompressibility of symmetric nuclear matter  and the nucleon effective mass at 
the empirical saturation density are respectively $K = 300$~MeV and   $M^{*} = 0.7 M$ 
(being $M$ the bare nucleon mass).
The inclusion of hyperons involves  new couplings, which can be written in terms of the 
nucleonic ones: 
$g_{\sigma Y} = x_{\sigma } g_{\sigma}$,~~ $g_{\omega Y} = x_{\omega } g_{\omega}$ 
and $g_{\rho Y} = x_{\rho } g_{\rho}$. 
In this model it is assumed that all the hyperons in the baryon octet have the same coupling. 
In this work we will consider $x_\sigma$ in the range of $0.6$ -- $0.8$.     
In addition, following Ref. \cite{gm}, we will take  $x_{\rho} = x_{\sigma}$, whereas 
the binding energy of the $\Lambda$ particle in symmetric nuclear matter, $B_\Lambda$,
\begin{equation}
 \left(\frac{B_\Lambda}{A}\right)=-28 \mbox{ MeV}= x_{\omega} \, g_{\omega}\, \omega_0-x_{\sigma}\,  g_{\sigma} \sigma 
\end{equation}
is used to determine $x_\omega$ in terms of $x_\sigma$.    
Notice that the case with $x_{\sigma} = 0.6$ produces stars with a larger hyperon population 
(for a given stellar gravitational mass) with respect to the case  $x_{\sigma} = 0.8$ \cite{gle,b6}.   
In addition to these two parametrizations for hyperonic matter (hereafter called $NY$ matter), we will consider the case of pure nucleonic matter (hereafter called  $N$ matter).

\section{Phase transition in beta-stable neutron star matter}
\label{phtr} 
The composition of neutron star matter is determined by the requirements of electric  
charge neutrality and equilibrium under the weak interaction processes ($\beta$-stable matter).   
Under such conditions, in the case of neutrino-free matter, the chemical potential $\mu_i$ 
of each particle species $i$ can be written in terms of two independent quantities, 
the baryonic and electric chemical potentials $\mu_b$ and $\mu_q$ respectively \cite{prak97},    
\be
\mu_i = b_i \mu_b - q_i \mu_q \ ,
\ee     
where $b_i$ is the baryon number of the species $i$, and $q_i$ denotes its charge in 
units of the electron charge magnitude. 
 
In the pure hadronic phase, $\mu_b = \mu_n$ is the neutron chemical potential,  and 
$\mu_q = \mu_e$ is the electron chemical potential. 
In the pure quark phase the quark chemical potentials $\mu_f$ ($f$=$u, d, s$) are 
related to  $\mu_b$ and $\mu_q$ by the formulas  
$\mu_u = (\mu_b - 2 \mu_q)/3$  and  $\mu_d = \mu_s = (\mu_b + \mu_q)/3$.

We next assume a first-order hadron-quark phase transition \cite{fodor} 
and, following Glendenning \cite{gle92}, we require global electric charge neutrality 
of bulk $\beta$-stable stellar matter. An important consequence of imposing global 
charge neutrality is that the hadronic and the quark phases can coexist for a finite 
range of pressures. This treatment of the phase transition is known in the 
literature as the Gibbs construction for the mixed phase. 
In this case the Gibbs conditions for phase equilibrium can be written  as 
\bea
\mu_{b,H} = \mu_{b,Q} \equiv \mu_b \\
\mu_{q,H} = \mu_{q,Q} \equiv \mu_q  \\
\label{g1}  
T_H = T_Q \equiv T \\ 
\label{g2} 
P_H(\mu_b, \mu_q, T) = P_Q(\mu_b, \mu_q, T)
\eea

With the purpose to compare with previous studies \cite{Baldo}, we also make use 
of the so-called Maxwell construction for the phase transition. 
In this case, one imposes that each phase in equilibrium is separately charge neutral. 
Now the conditions for phase equilibrium can be written  as 
\bea
\mu_{b,H} = \mu_{b,Q} \equiv \mu_b \\
\label{max1}  
T_H = T_Q \equiv T \\ 
\label{max2} 
P_H(\mu_{b}, \mu_{q}(\mu_b), T) = P_Q(\mu_b, \mu^\prime_q(\mu_b), T) \ . 
\eea 
In this case the electric chemical potential  $\mu_q = \mu_e$ has a discontinuity \cite{gle92} 
at the interface between the two phases
{\footnote{For a detailed discussion of the differences between the Maxwell and the Gibbs 
phase transition constructions including the effects of surface tension and electromagnetic 
energy see Ref. \cite{dima03,mar07}.}}. 

In the following we consider the case of cold (i.e. $T = 0$) matter, which is appropriate  
to describe neutron stars interiors at times larger than about a few minutes after 
their formation \cite{prak97}.  

\begin{figure}
\centering
\includegraphics[scale=0.35]{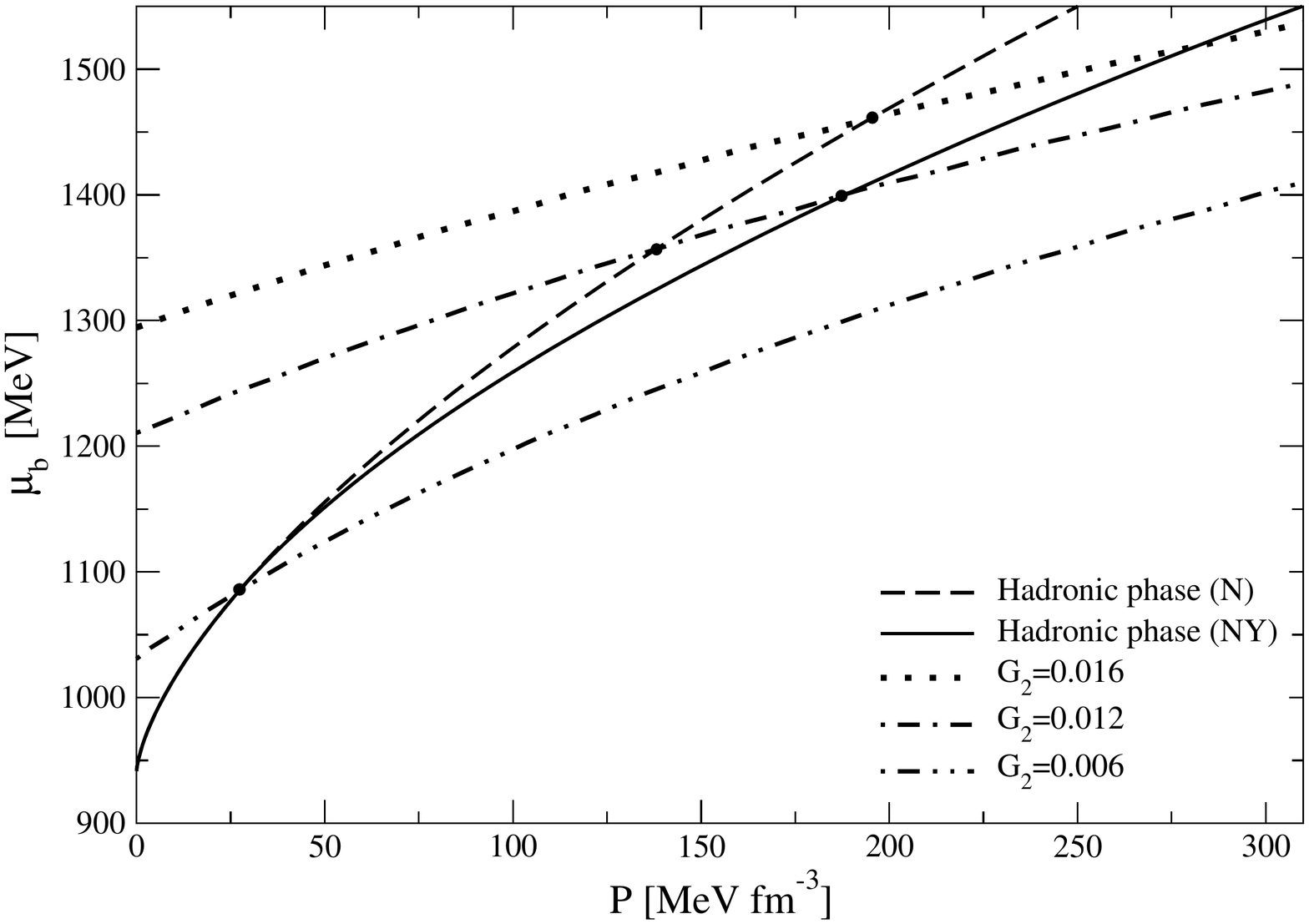}
\caption{Baryon chemical potential $\mu_b$ versus pressure $P$ in cold ($T=0$) $\beta$-stable matter. 
Curves for the quark phase are relative to three different values of the gluon condensate 
$G_{2}$ reported in GeV$^4$, and  $V_1 = 0.01 \, {\rm GeV}$.     
Curves for the hadronic phase are relative to hyperonic matter (NY) with $x_{\sigma} = 0.7$ 
and to pure nucleonic matter (N).}   
\label{fig1}
\end{figure}
\begin{figure}
\centering
\includegraphics[scale=0.35]{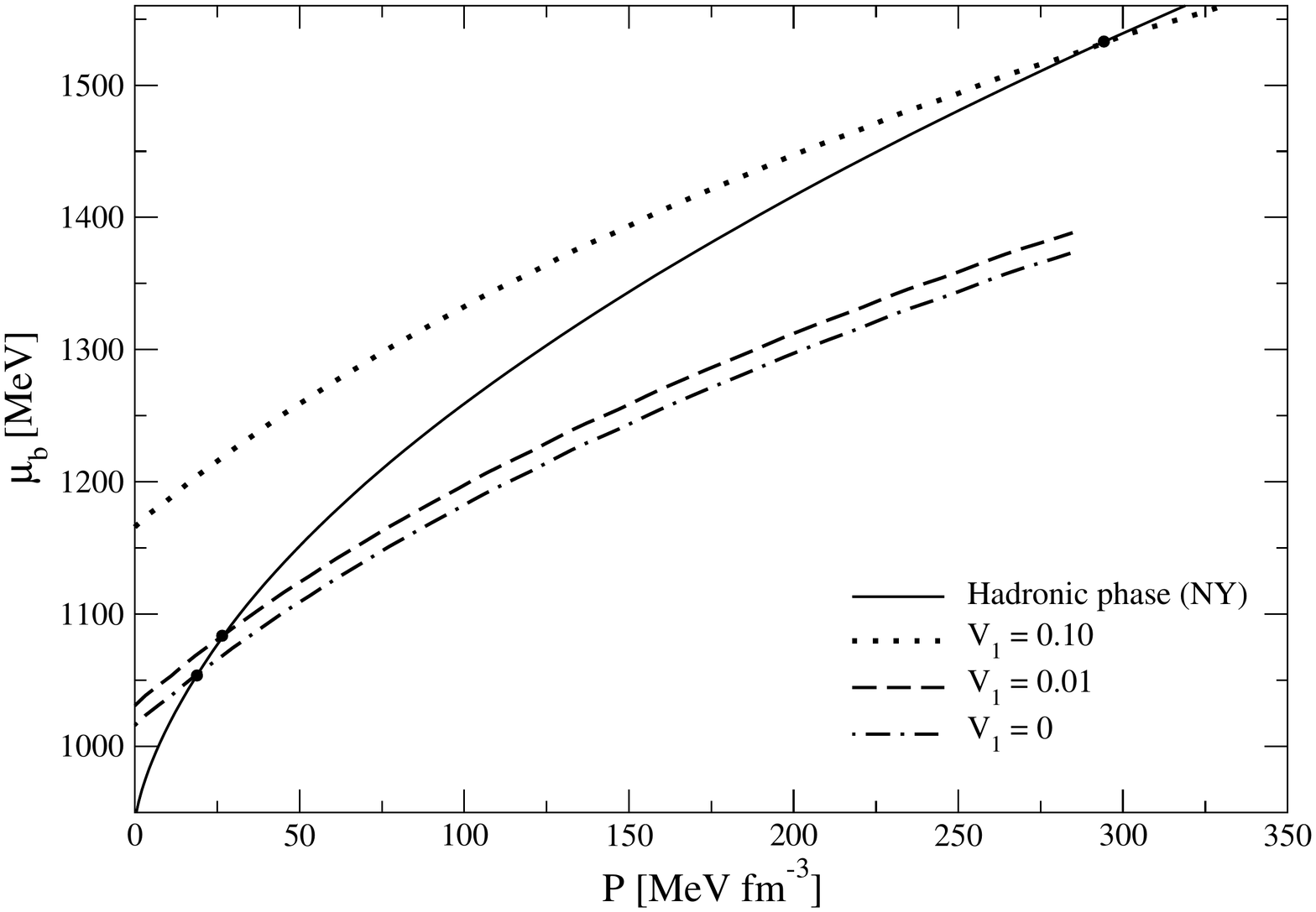} 
\caption{Baryon chemical potential $\mu_b$ versus pressure $P$ in cold ($T=0$) $\beta$-stable matter. 
Curves for the quark phase are relative to three different values of the large-distance static 
$Q\bar{Q}$ potential $V_1$ reported in GeV, and for $G_{2} = 0.006 \, {\rm GeV}^4$. 
The curve for the hadronic phase is relative to hyperonic matter (NY) with $x_{\sigma} = 0.6$.}   
\label{fig2}
\end{figure} 

In Fig.~\ref{fig1} we plot the relation between the baryon chemical potential $\mu_b$ 
and the total (i.e. baryonic plus leptonic contributions) pressure $P$, 
in $\beta$-stable matter, for the hadron and the quark phases in the case 
of the Maxwell construction. 
For the hadronic phase we consider hyperonic matter (NY)  with $x_\sigma = 0.7$ (continuous line) and pure nucleonic matter (N) (dashed line). For the quark phase we use three different values of the gluon 
condensate  $G_2 = 0.006, \,  0.012, \, 0.016$~GeV$^4$ and a 
common value $V_1 = 0.01 \, {\rm GeV}$ for the large-distance static $Q\bar{Q}$ potential. 
The phase transition occurs at the intersection point between the curves describing the two different phases.  
This crossing point is significantly affected by the value of the gluon condensate, in particular  
when $G_{2}$ increases the onset of the deconfinement transition is shifted to higher pressure 
(higher baryon chemical potential).
It is worthwhile to note also that the presence of hyperons in the hadronic phase moves 
the phase transition point to larger pressures.   
Similar results have been found using different values for the hyperon-nucleon 
couplings  ($x_\sigma = 0.6, 0.8$)  for the hadronic EOS.  

Keeping a fixed value of the gluon condensate, $G_{2} = 0.006 \, {\rm GeV}^4$, we show in 
Fig.~\ref{fig2} the effects of $V_{1}$ on the phase transition point. 
As one can see, only slight differences exist between the cases $V_{1} = 0$ and 
$V_{1} = 0.01\, {\rm GeV}$ while for $V_{1} = 0.1 \, {\rm GeV}$ the transition point 
is shifted to a very high value of the pressure.  
This should be expected just looking at Eqs.~\ref{P_FG0} and \ref{mu_q}, where it turns 
out clearly that pressure is a decreasing function of $V_{1}$. 

\begin{figure}
\centering
\includegraphics[scale=0.35]{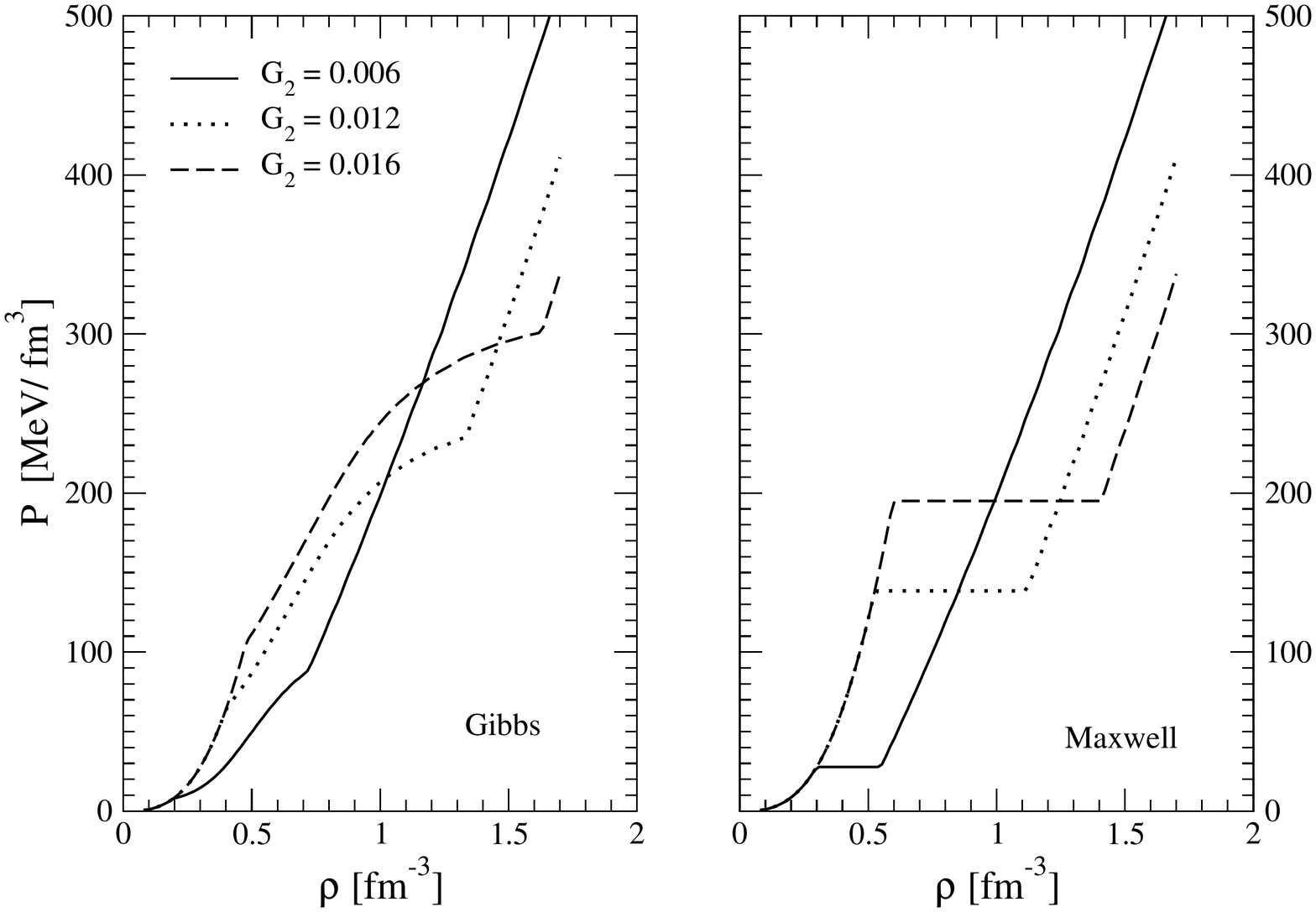} 
\caption{Total pressure $P$ of cold $\beta$-stable matter as a function of the baryon number 
density $\rho$, for different values of the gluon condensate $G_{2}$ (reported in GeV$^4$ units)   
and for $V_{1} = 0.01 \, {\rm GeV}$  in the case of the Gibbs (left panel) 
and Maxwell construction (right panel) for the phase transition.  
Results are relative to pure nucleonic matter (N) for the hadronic phase.}
\label{P-rho_N}
\end{figure} 
\begin{figure}
\centering
\includegraphics[scale=0.35]{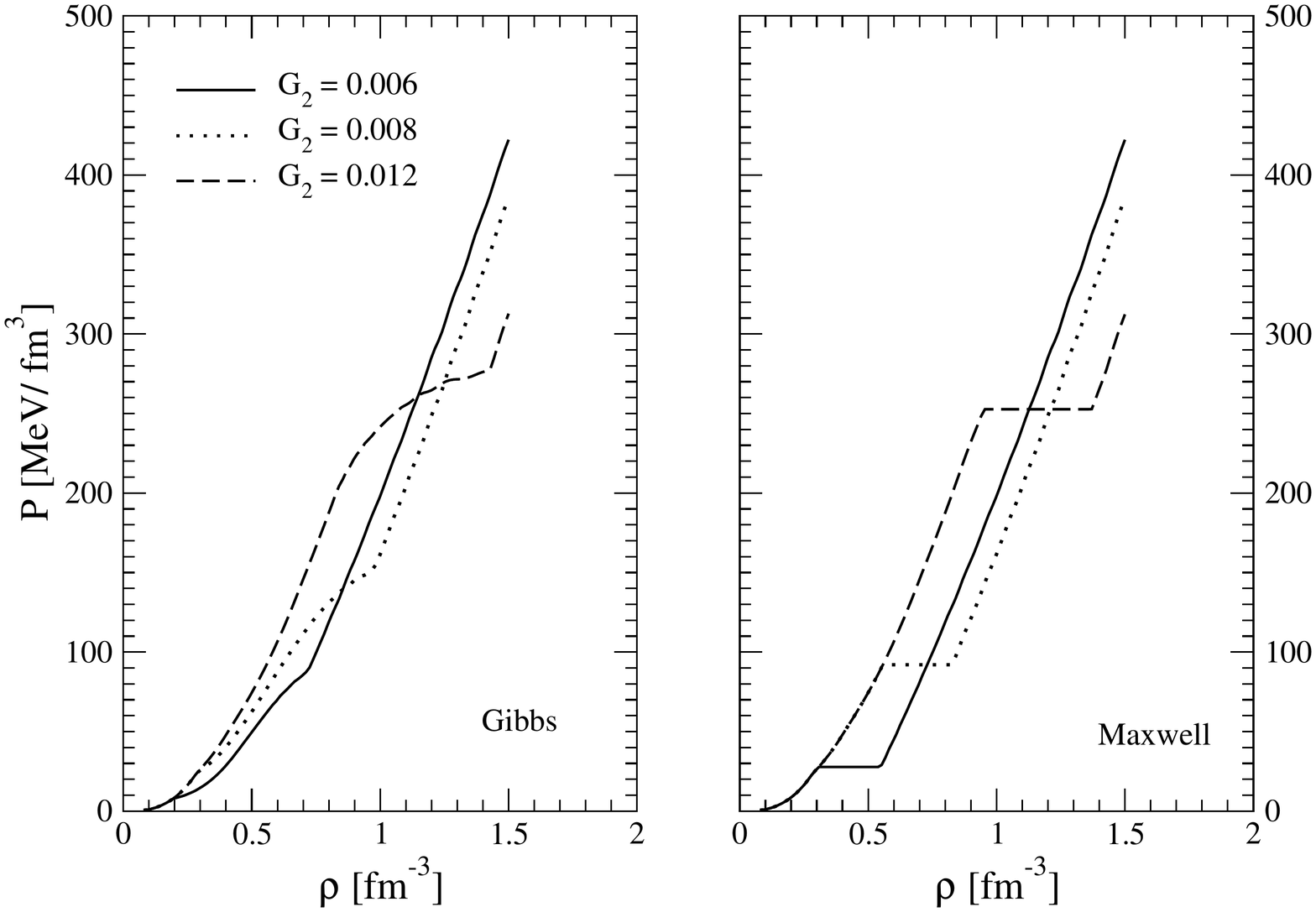}
\caption{Same as the previous figure, but with hyperonic matter (NY) with $x_{\sigma} = 0.6$ 
for the hadronic phase.} 
\label{P-rho_NY}
\end{figure} 

In Figs. \ref{P-rho_N} and \ref{P-rho_NY}, we show the pressure for 
$\beta$-stable matter as a function of the  baryon number density $\rho$ in the 
case of pure nucleonic matter (Fig. \ref{P-rho_N}) and hyperonic matter 
with $x_\sigma = 0.6$ (Fig. \ref{P-rho_NY}) for the hadronic phase. 
For the quark phase we consider different values of the gluon condensate $G_{2}$. 
Results on the left (right) panels of these two figures refer to the Gibbs (Maxwell) 
construction for the phase transition. 
As expected, the Maxwell construction corresponds to a constant pressure in the 
baryon number density range of the coexistence region, 
whereas in the Gibbs construction the pressure increases monotonically  
with $\rho$ \cite{gle92,gle,bhat10}. 
Increasing the value of $G_{2}$ causes a shift of the phase transition to larger baryon 
densities for both the Gibbs and Maxwell constructions.

\section{Neutron star structure} 
\label{NSstr}

In this section we show the results of our calculations of hybrid stars structure.  
To this purpose we integrate the well-known Tolman, Oppenheimer and Volkov
relativistic hydrostatic equilibrium equations (see e.g. \cite{gle,pawel}) to get 
various stellar properties for a fixed EOS.  

We report the results of a systematic study in which we vary the value of the gluon  
condensate $G_{2}$ between the constraints imposed by QCD sum rules \cite{QCDsum-rules}.   
To model the neutron star crust we have used the EOS of Ref. \cite{crust}.  

Unless otherwise specified, all the results presented in the following have been 
obtained using the Gibbs construction to model the hadron-quark phase transition 
and taking $V_{1} = 0.01 \, {\rm GeV}$. 

\begin{figure}
\centering
\includegraphics[scale=0.35]{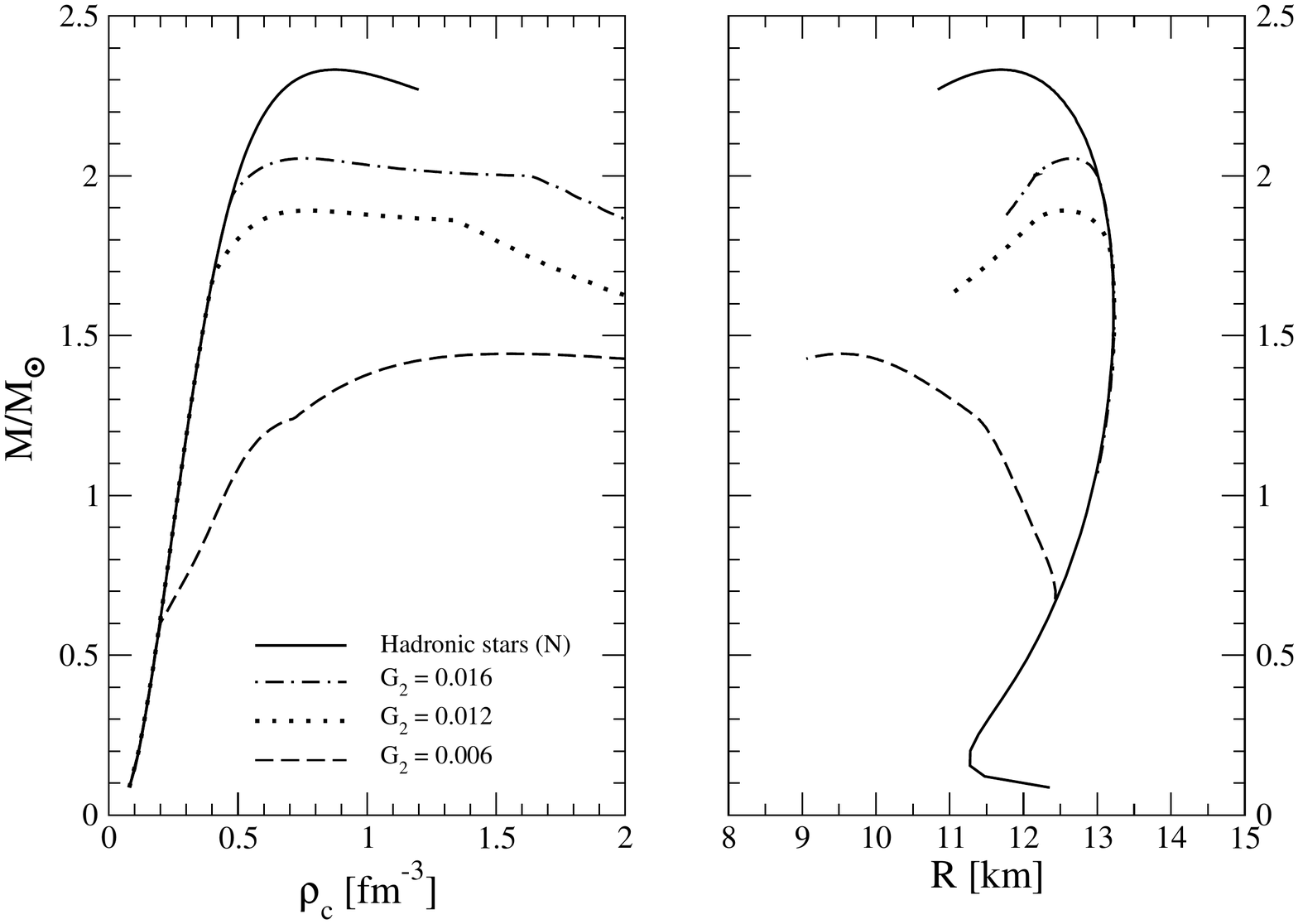}
\caption{Stellar gravitational mass $M$ versus central baryon number density 
$\rho_c$ (left panel) and versus stellar radius $R$ (right panel) for hybrid stars for 
several values of the gluon condensate $G_{2}$ (reported in GeV$^4$ units) 
and for $V_{1} = 0.01 \, {\rm GeV}$.   
The continuous line in both panels refers to the pure hadronic stars (i.e. compact 
stars with no quark matter content).  
For the EOS of hadronic phase only nucleons (N) are included.}
\label{Mrho+MR_N}
\end{figure}

Let us first consider the case in which the hadronic phase does not contain hyperons, 
{\it i.e.} the case of pure nucleonic matter (N).   
In Fig.~\ref{Mrho+MR_N} we report the stellar gravitational mass $M$ 
(in unit of the solar mass $M_\odot = 1.99 \times 10^{33} \rm{g}$) 
versus the central baryon number  density $\rho_{c}$ (left panel) and the 
mass versus radius $R$ (right panel) in the case of pure nucleonic stars (continuous line)  
and of hybrid stars for different $G_{2}$.  
We obtain stable hybrid star configurations for all the considered values of the gluon 
condensate, with maximum masses ranging from  
$M_{max} = 1.44 \, M_\odot$ (case with $G_{2} = 0.006 \, {\rm GeV}^4$) to  
$M_{max} = 2.05 \, M_\odot$ ($G_{2} = 0.0016 \, {\rm GeV}^4$).   
Notice that the hybrid star branch of the stellar equilibrium configurations shrinks as $G_{2}$ 
is increased.  This is in full agreement with the results for the EOS reported 
in the left panel of Fig. \ref{P-rho_N}.  
This behavior is different with respect to the one found by the authors of Ref. \cite{Baldo} 
where the stability window of hybrid star configurations for a nucleonic equation of state 
was restricted between $0.006 \, {\rm GeV}^4 < G_{2} < 0.007 \, {\rm GeV}^4$, 
while no hadron-quark phase transition occurred once hyperons were included in the EOS.\\   

\begin{figure}
\centering
\includegraphics[scale=0.35]{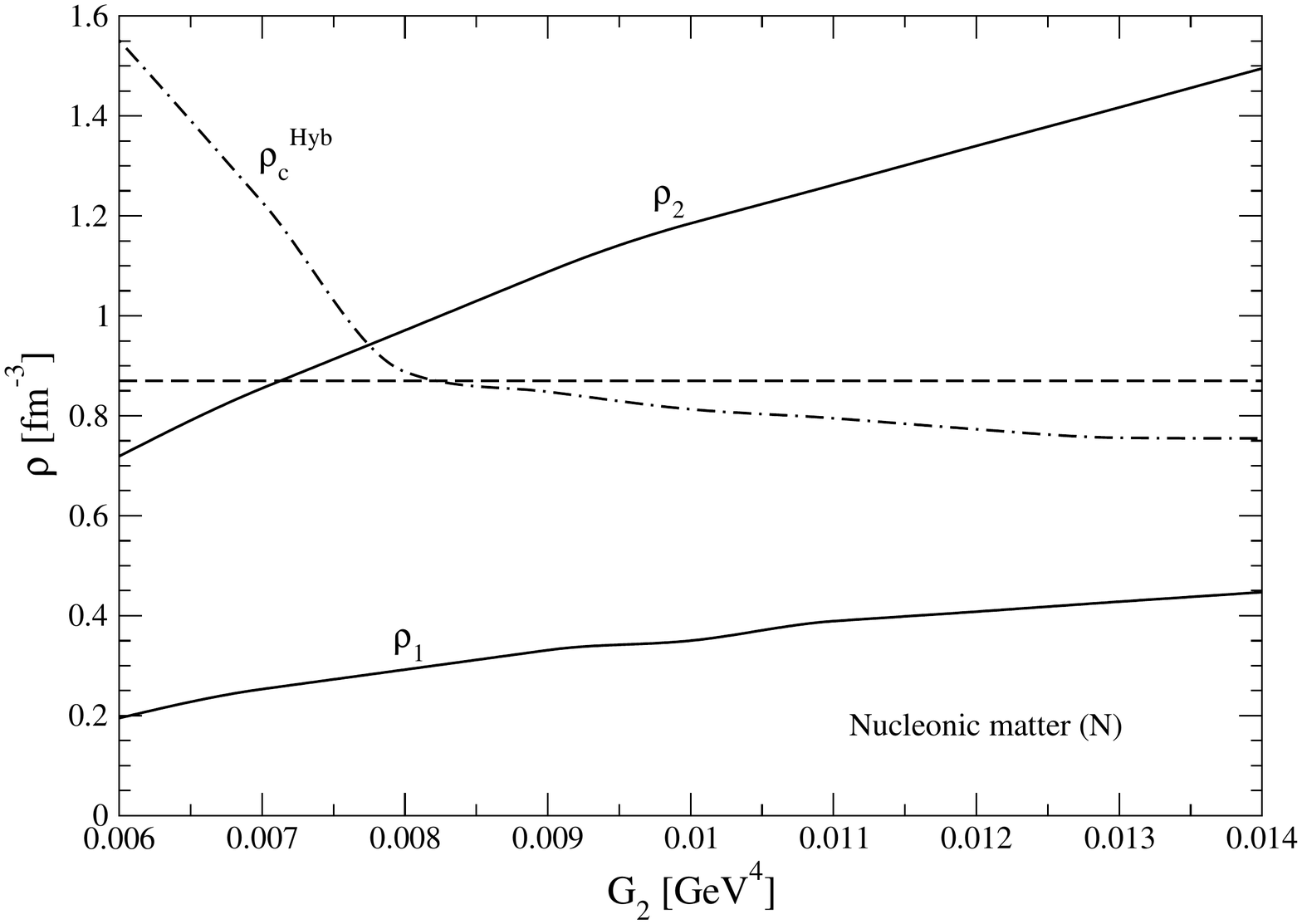}
\caption{Quark-hadron phase transition boundaries in $\beta$-stable matter as a  
function of the gluon condensate $G_2$ and for $V_{1} = 0.01 \, {\rm GeV}$. 
The onset of quark-hadron mixed phase occurs at the baryon number density $\rho_1$, 
and the pure quark phase begins at $\rho_2$.  Also shown is the central baryon number 
density $\rho_c^{Hyb}$ of the maximum mass hybrid star. The horizontal dashed line represents 
the value of the central baryon number density $\rho_c^{HS}$ of the maximum mass 
pure hadronic star.  
For the EOS of hadronic phase only nucleons (N) are included. }
\label{rho_g2_N}
\end{figure} 

\begin{figure}
\centering
\includegraphics[scale=0.35]{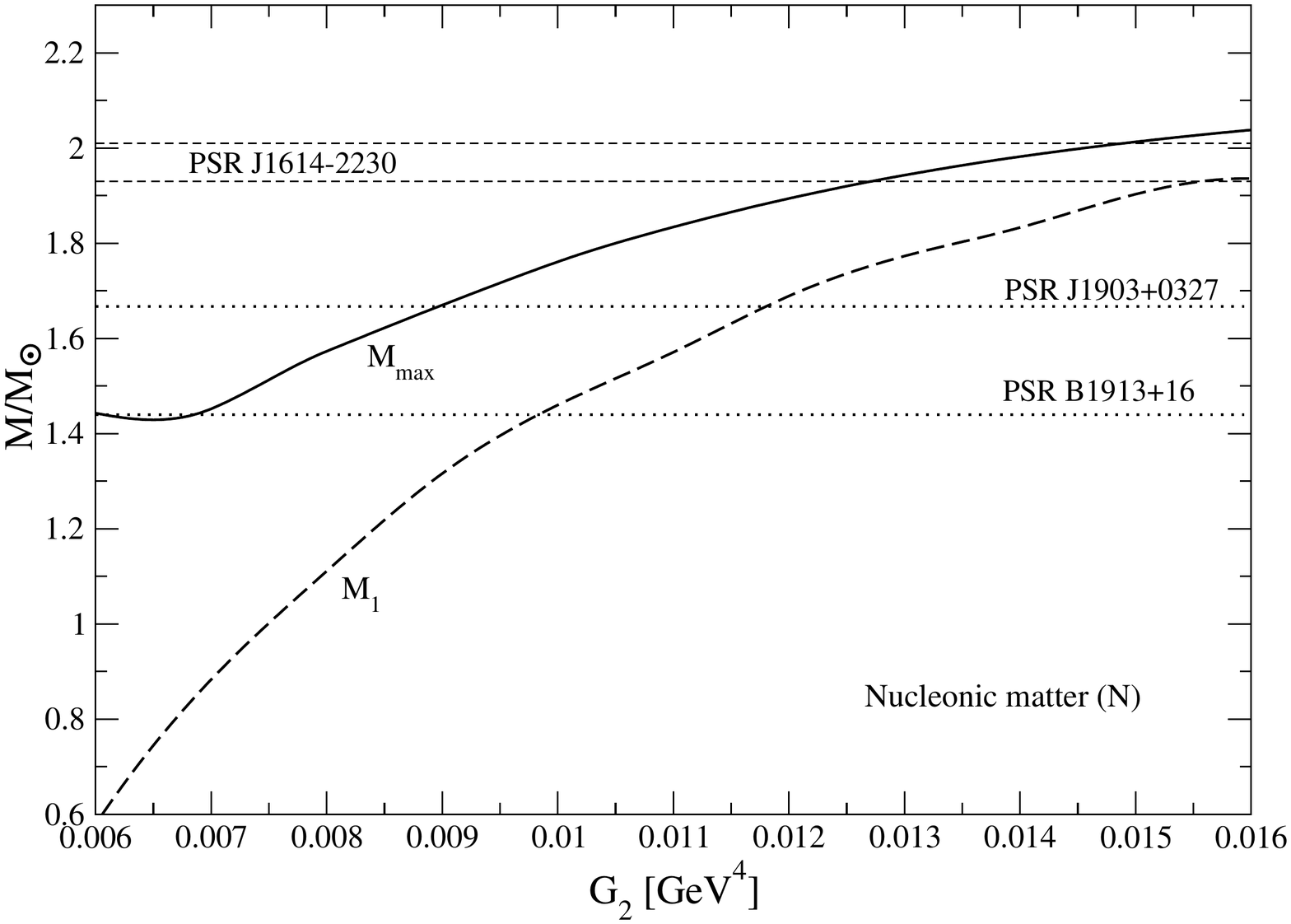}
\caption{Gravitational maximum mass for hybrid stars (continuous line) and gravitational 
mass $M_1$ (dashed line) of the star with central baryon number density $\rho_1$ corresponding 
to the onset of mixed quark-hadron phase as a function of the gluon condensate $G_{2}$ 
and for $V_{1} = 0.01 \, {\rm GeV}$.   
For the EOS of hadronic phase only nucleons (N) are included.} 
\label{Mmax_g2_N}
\end{figure} 

In Fig. \ref{rho_g2_N} we plot the quark-hadron phase transition boundaries in 
$\beta$-stable matter as a function of  $G_2$ in the case in which the 
hadronic phase does not contain hyperons (pure nucleonic matter).  
The onset of the deconfinement transition (i.e. the onset of the 
quark-hadron mixed phase) occurs at the baryon number density $\rho_1$, 
and the pure quark phase begins at $\rho_2$.    
Also shown is the central baryon number density $\rho_c^{Hyb}$ of the 
maximum mass hybrid star (dotted-dashed line). 
Stable hybrid star configurations have central densities $\rho_c$ located in the region 
of the  $\rho$--$G_2$ plane between the dotted-dashed line and the lower continuous line, 
i.e. $\rho_1 < \rho_c  \leqslant \rho_c^{Hyb}$.   
Notice that $\rho_c^{Hyb} > \rho_2$ when the gluon condensate is in the range 
$0.006  \, \rm{GeV}^4 < G_{2}  \leqslant 0.0077  \, \rm{GeV}^4$.  
For these $G_{2}$ values all hybrid stars with a central density in the range 
$\rho_2 < \rho_c \leqslant  \rho_c^{Hyb}$  possess a pure quark matter core.  
Finally the horizontal dashed line represents the value of the central baryon number  
density $\rho_c^{HS}$ of the maximum mass pure nucleonic star.   

In Fig. \ref{Mmax_g2_N} we draw the maximum mass $M_{max}$ for hybrid stars 
(continuous line) and the mass $M_1 = M(\rho_1)$ (dashed line) of the star with central 
baryon number density $\rho_1$ corresponding to the onset of the mixed phase.    
These two quantities are plotted  as a function of the gluon condensate $G_{2}$ 
for the case in which the hadronic phase does not contain hyperons.  
Stable hybrid star configurations correspond to the region of the $M$--$G_2$ plane 
between the continuous and the dashed line.    
Stellar configurations in the region below the dashed line $M_1$ do not possess any 
deconfined quark matter in their center (nucleonic stars).    

To compare our results with measured neutron star masses, we report in 
the same Fig. \ref{Mmax_g2_N} the values of the masses of the following pulsars: 
PSR~B1913+16  with $M = 1.4398 \pm 0.0002 \, M_\odot$ \cite{HT75},     
PSR~J1903+0327 with $M = 1.667 \pm 0.021 \, M_\odot$ \cite{frei11} and  
PSR~J1614-2230  with $M = 1.97 \pm 0.04 \, M_\odot$ \cite{demo10}.  \\  
The mass of PSR~J1614-2230 gives the strongest constraint on the possible value 
of the gluon condensate.  In fact, using the lower bound of the measured mass of 
PSR~J1614-2230, we get  $G_{2} \gtrsim 0.0129~{\rm GeV}^4$.
Thus for values of the gluon condensate in the range  
$0.0129~{\rm GeV}^4  \lesssim G_{2} \leqslant G_{2}^{*} \simeq 0.018~{\rm GeV}^4$,    
PSR~J1614-2230 is a hybrid star,  whereas PSR~B1913+16 and PSR~J1903+0327 
are pure nucleonic stars. In the above specified range, $G_{2}^{*}$ is defined 
by the condition $M_1(G_{2}^{*}) = 2.01~M_\odot$, the upper bound of the 
measured mass of PSR~J1614-2230.  
Thus for $G_{2} > G_{2}^{*}$ PSR~J1614-2230 is a pure nucleonic star.  

\begin{figure}
\centering
\includegraphics[scale=0.35]{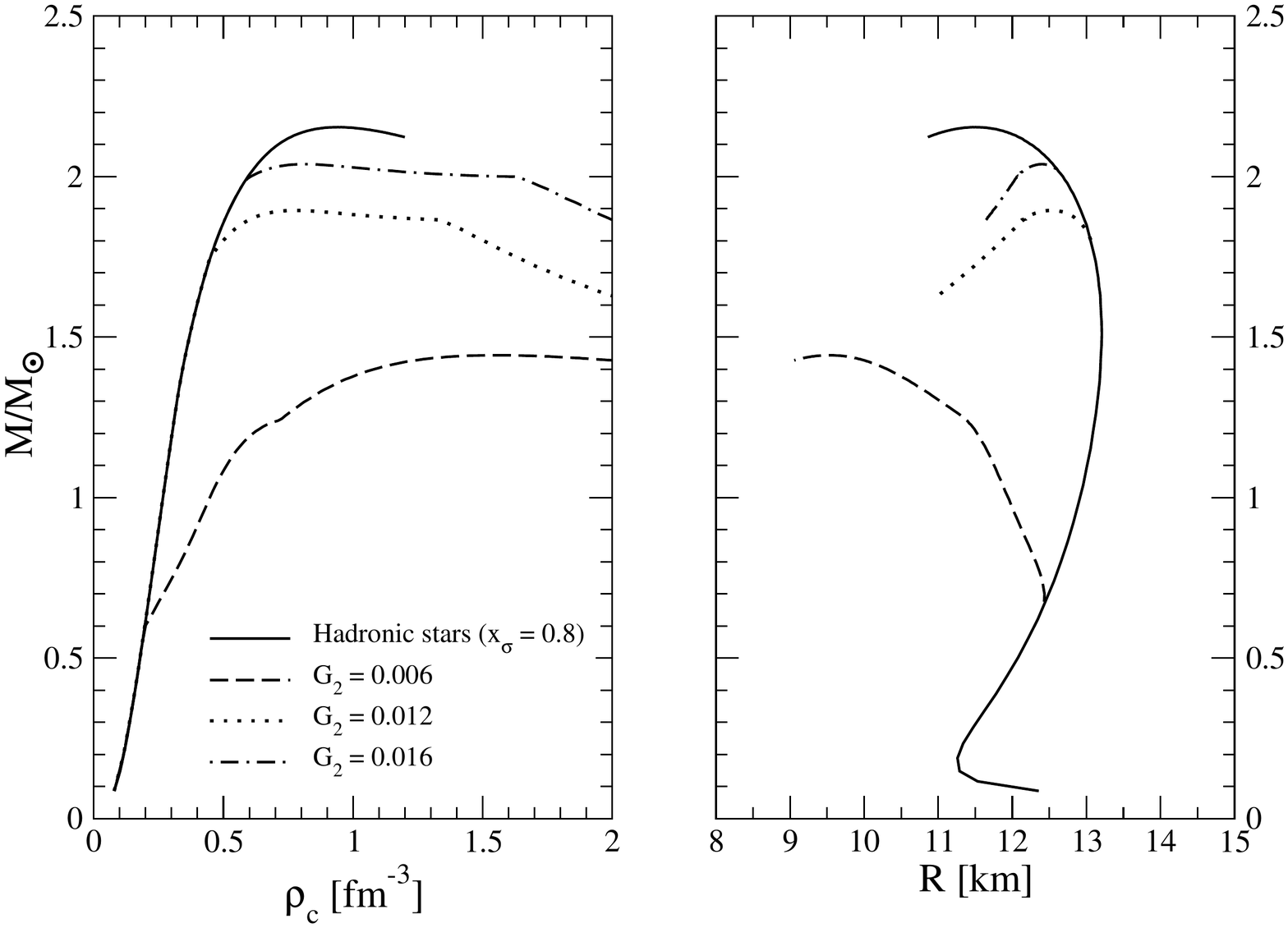}
\caption{Stellar gravitational mass $M$ versus central number density $\rho_c$ (left panel) 
and versus stellar radius $R$ (right panel) for hybrid stars for several values of the gluon 
condensate $G_{2}$ (reported in GeV$^4$ units) and for $V_{1} = 0.01 \, {\rm GeV}$.   
The continuous line in both panels refers to the pure hadronic star  sequence. 
The hadronic phase consists of hyperonic matter (NY) and is described by the GM1 model 
with $x_\sigma = 0.8$.}   
\label{Mrho+MR_xs08}
\end{figure} 

\begin{figure}
\centering
\includegraphics[scale=0.35]{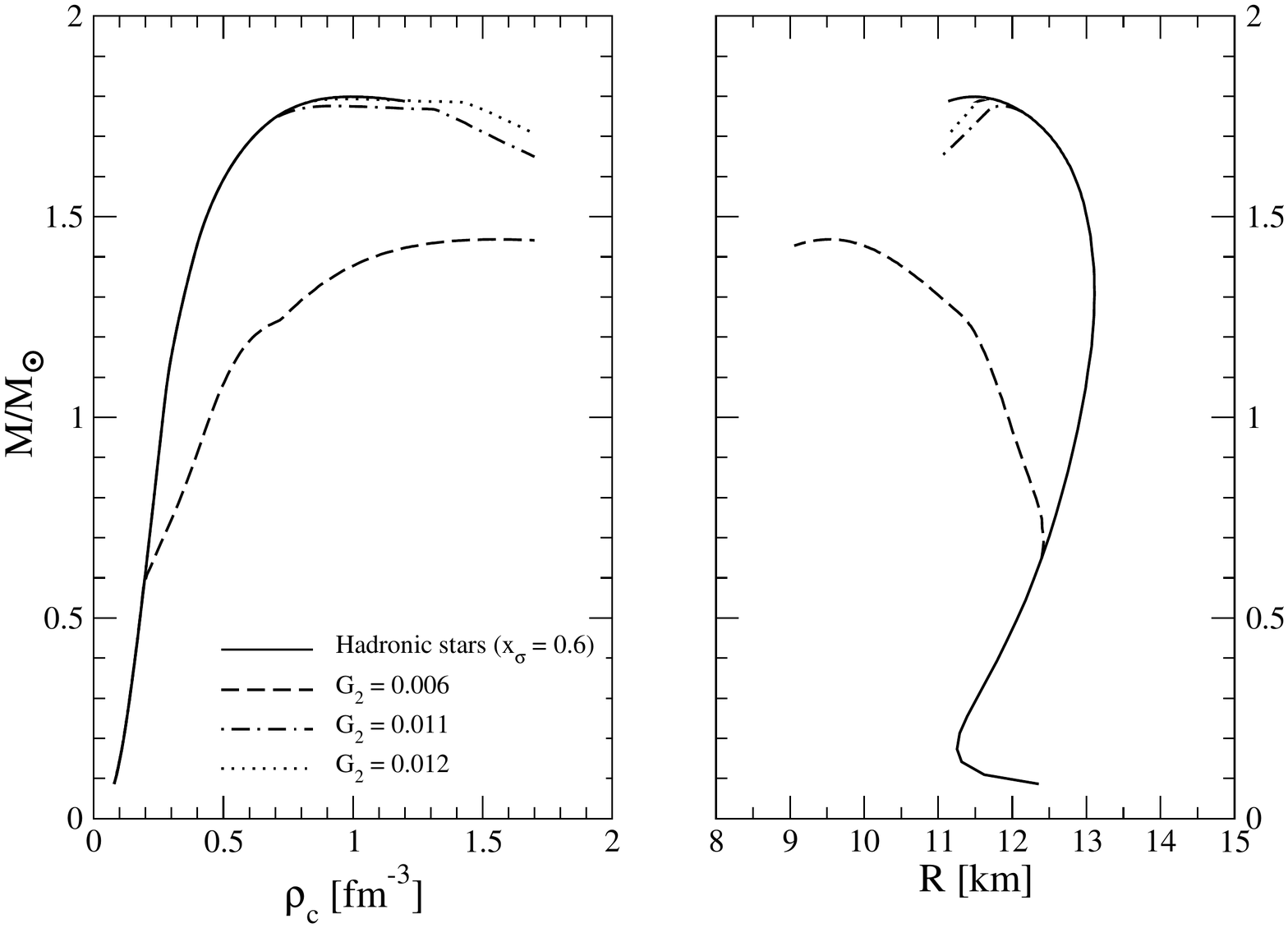}
\caption{Same as Fig. \ref{Mrho+MR_xs08} but with $x_\sigma = 0.6$.}  
\label{Mrho+MR_xs06}
\end{figure} 

\begin{table} 
\begin{center}
\bigskip
\begin{ruledtabular}
\begin{tabular}{l|ccccccc}
  $x_{\sigma}$   & $G_2$  & $M_{max}$  & $\rho_c^{Hyb}$  & $R$ & $M_{max}^{HS}$ & $\rho_{c}^{HS}$ & $R^{HS}$ \\
\hline
                       & 0.006        & 1.44      & 1.55 &   9.54     &        &         &       \\
  N                   & 0.012        & 1.89      & 0.77 &  12.55    & 2.33 &  0.87 & 11.70 \\
                       & 0.016        & 2.05      & 0.75 &  12.66    &        &         &       \\                     
\hline

                       & 0.006        & 1.44      & 1.56 &    9.52    &        &        &       \\
 0.8                 & 0.012        & 1.89      & 0.77 &  12.53    & 2.15 &  0.94  & 11.50 \\ 
                       & 0.016        & 2.04      & 0.81 &  12.40    &        &        &       \\

\hline
                       & 0.006        & 1.43      & 1.56 &   9.51    &       &        &         \\
 0.6                 & 0.010        & 1.73      & 0.89 &  12.00    & 1.80  &  1.00  & 11.49   \\
                      & 0.013        & 1.80      & 1.00 &  11.49    &       &        &         \\
\end{tabular}
\end{ruledtabular}
\end{center} 
\caption{Properties of the maximum mass configuration for hybrid stars for 
different values of the gluon condensate $G_{2}$ in GeV$^{4}$ (second column)  
and for $V_{1} = 0.01 \, {\rm GeV}$. 
The parameter $x_\sigma$ (first column) fixes the hyperons coupling constants as described 
in Sec. \ref{eoshm}. The entry $N$ in the first column refers to the case of pure nucleonic matter 
for the hadronic phase.  
For the case $x_{\sigma} = 0.6$, hybrid stars are possible only for 
$G_2  \lesssim 0.013 \, {\rm GeV}^4$ (see Fig. \ref{rho_g2_xs06}).     
The mass $M_{max}$, the central baryon number density $\rho_c^{Hyb}$ and the radius $R$ of the maximum mass configuration are reported respectively in the third, fourth and fifth columns. 
The quantities with the label HS refer to the case of purely hadronic stars.    
Stellar masses are reported in units of the solar mass $M_\odot = 1.99 \times 10^{33} \rm{g}$, 
central densities are given in $\rm{fm}^{-3}$, and stellar radii in km.}  
\label{t:mass}
\end{table} 

\begin{figure}
\centering
\includegraphics[scale=0.35]{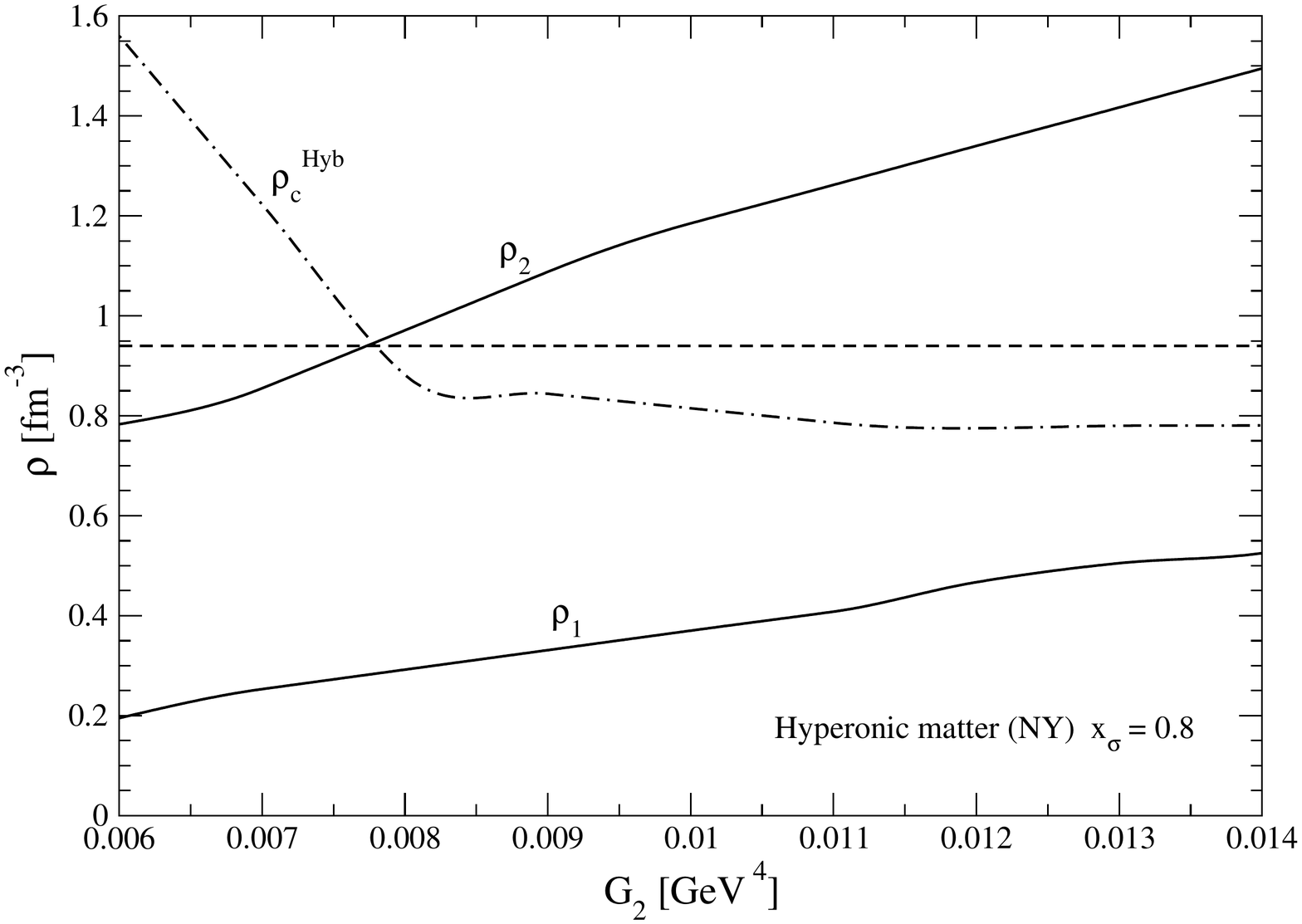}
\caption{Quark-hadron phase transition boundaries in $\beta$-stable matter as a  
function of the gluon condensate $G_2$ and for $V_{1} = 0.01 \, {\rm GeV}$.  
The onset of quark-hadron mixed phase occurs at the density $\rho_1$, 
and the pure quark phase begins at $\rho_2$.  Also shown is the central number density 
$\rho_c^{Hyb}$ of the maximum mass hybrid star. The horizontal dashed line represents 
the value of the central density $\rho_c^{HS}$ of the maximum mass pure hadronic star.  
For the EOS of hadronic phase only nucleons and hyperons (NY) are included with $x_\sigma = 0.8$.}
\label{rho_g2_xs08}
\end{figure} 
\begin{figure}
\centering
\includegraphics[scale=0.35]{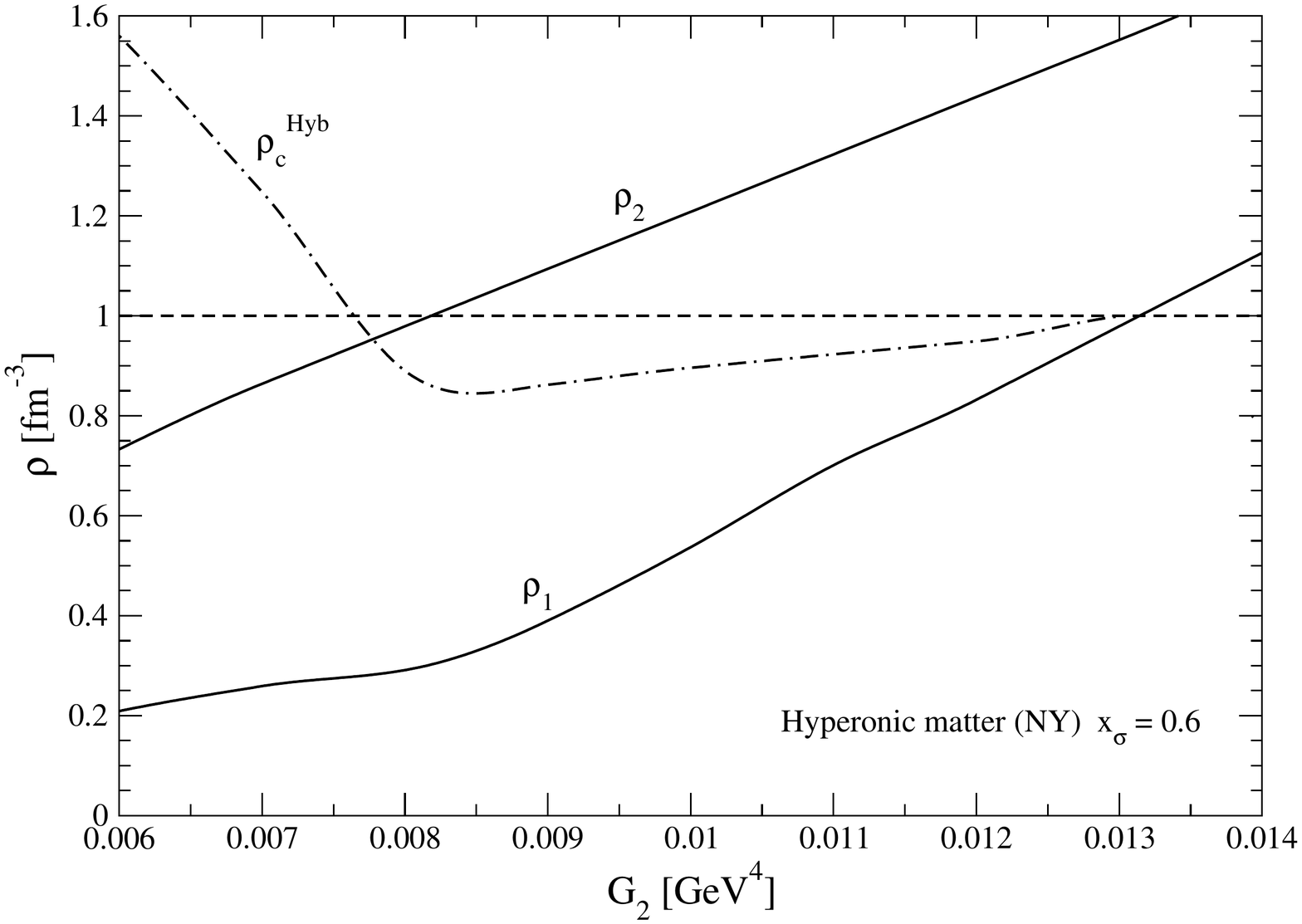}
\caption{Same as Fig. \ref{rho_g2_xs08} but with $x_\sigma = 0.6$.} 
\label{rho_g2_xs06}
\end{figure} 

\begin{figure}
\centering
\includegraphics[scale=0.35]{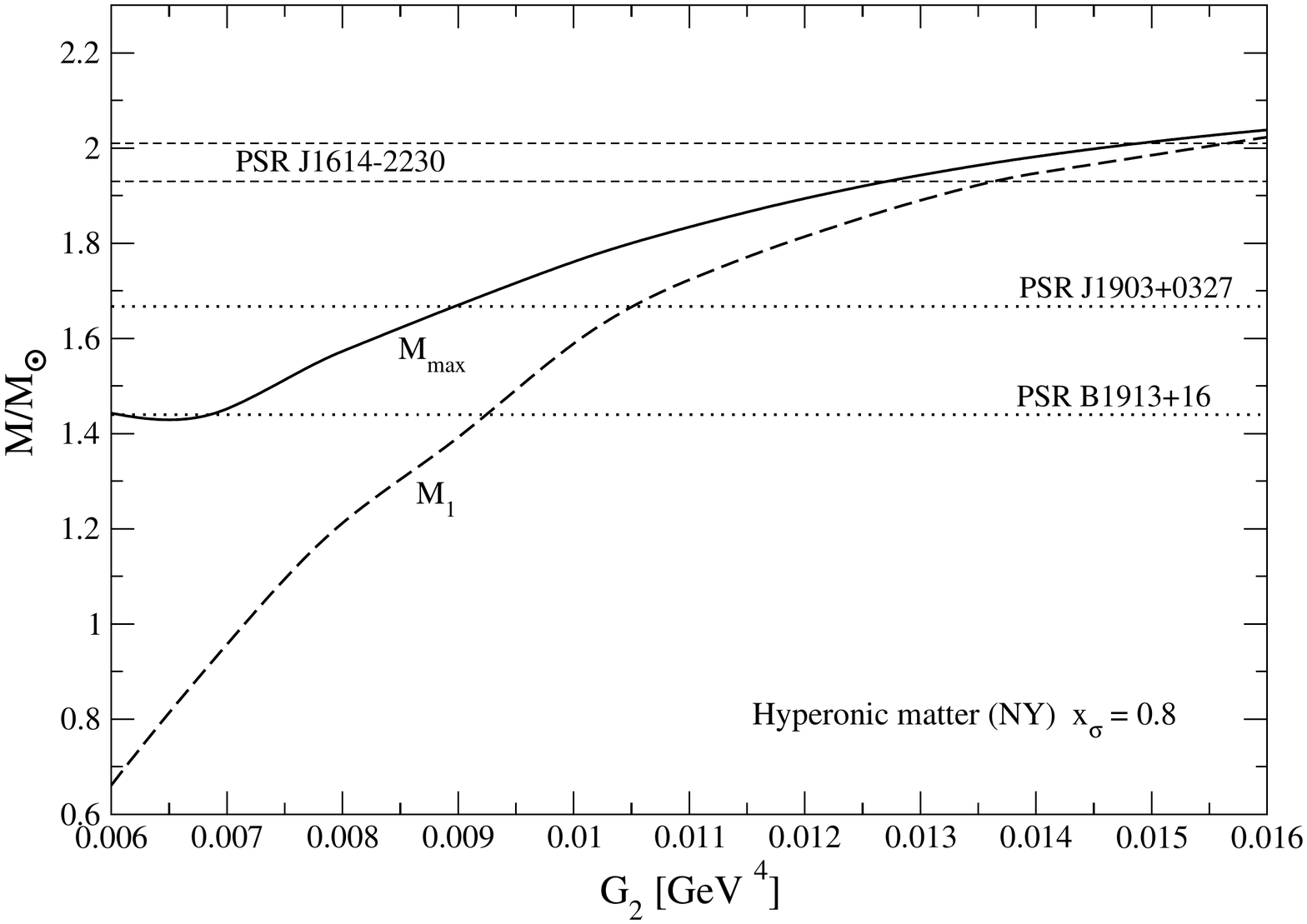}
\caption{Gravitational maximum mass for hybrid stars (continuous line) and gravitational 
mass $M_1$ (dashed line) of the star with central density $\rho_1$ corresponding to the 
onset of mixed quark-hadron phase as a function of the gluon condensate $G_{2}$ 
and for $V_{1} = 0.01 \, {\rm GeV}$.   
The hadronic phase consists of hyperonic matter (NY) and is described 
by the GM1 model with $x_\sigma = 0.8$.}  
\label{Mmax_g2_xs08}
\end{figure} 
\begin{figure}
\centering
\includegraphics[scale=0.35]{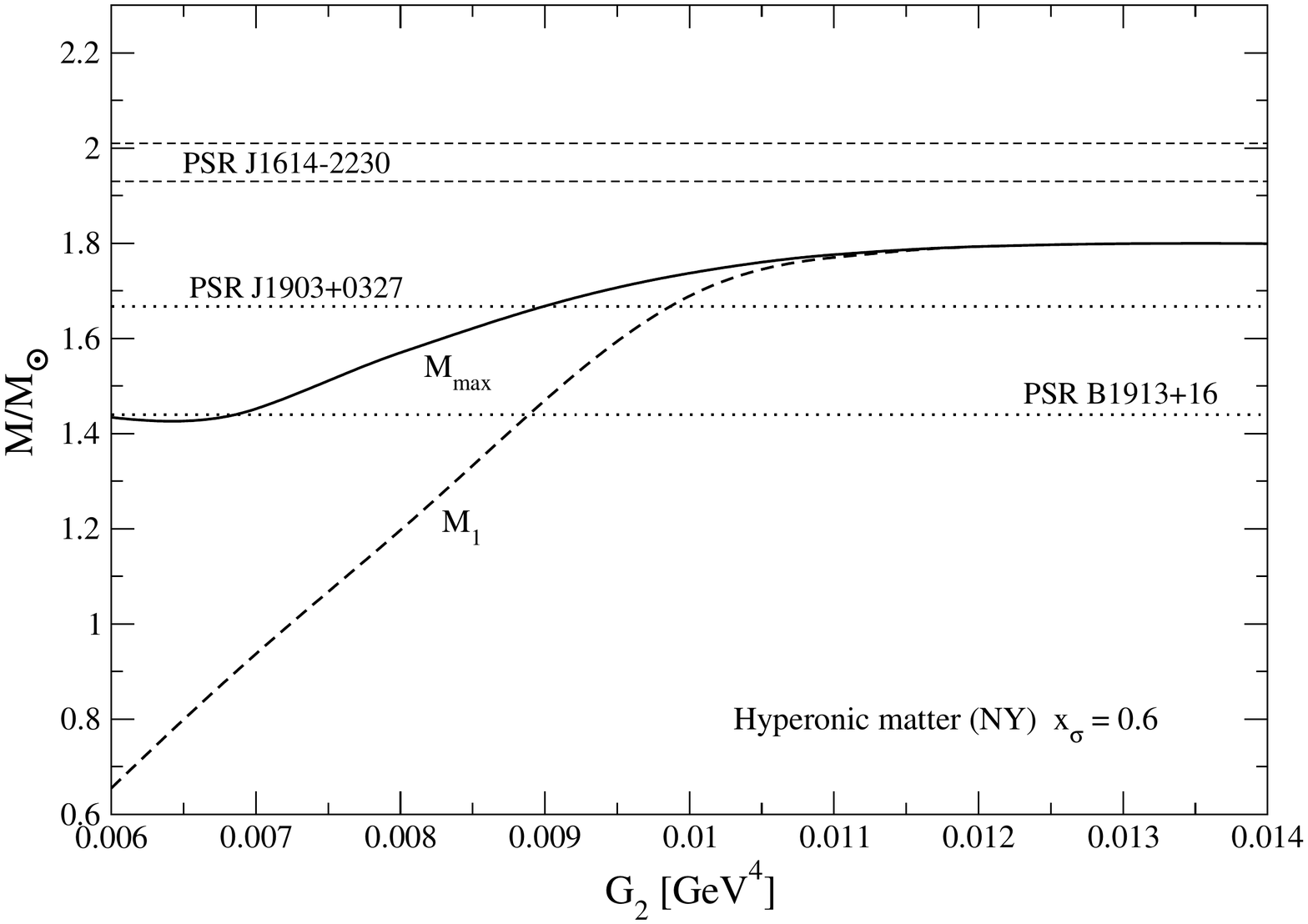}
\caption{Same as Fig. \ref{Mmax_g2_xs08} but with $x_\sigma = 0.6$.} 
\label{Mmax_g2_xs06}
\end{figure} 

We now consider the case in which the hadronic phase contains hyperons (NY matter), 
and we show the results of calculations for two different sets of the hyperon 
couplings: one corresponding to  $x_{\sigma} = 0.8$ and the other to  $x_{\sigma}=0.6$. 

In Fig. \ref{Mrho+MR_xs08} we plot for the case $x_{\sigma} = 0.8$ the stellar gravitational 
mass $M$ versus the central baryon number density $\rho_{c}$ (left panel) 
and versus the stellar radius $R$ (right panel), in the case of hyperonic stars (continuous line) 
and of hybrid stars for different $G_{2}$ values.  
The same quantities are depicted in Fig. \ref{Mrho+MR_xs06} for the case $x_{\sigma}=0.6$. 
As it is well known \cite{gle85}, the presence of hyperons reduces the value of the 
maximum mass of pure hadronic star (i.e. compact stars with no quark matter content) 
from $M_{max} = 2.33 \, M_\odot$ in the case of pure nucleonic stars (continuous line in 
Fig. \ref{Mrho+MR_N}) to $M_{max} = 1.80 \, M_\odot$ in the case of hyperonic stars with 
$x_{\sigma}=0.6$ (continuous line in Fig. \ref{Mrho+MR_xs06}). 
More interesting is to evaluate the effect of hyperons on the hybrid star sequence 
as a function of the gluon condensate.  As we can see, comparing the results in 
Figs. \ref{Mrho+MR_xs08} and \ref{Mrho+MR_xs06} with those in Fig. \ref{Mrho+MR_N}, 
for  ``low'' values of the gluon condensate (i.e. $G_2  \lesssim 0.008 \, {\rm GeV}^4$),
the hybrid star sequence is unaffected by the presence of hyperons.  
In fact, for these low values of $G_2$ the threshold density for hyperons is larger than 
the density $\rho_1$ for the onset of the quark-hadron mixed phase (see Fig. \ref{fig1}). \\  
As in the case of pure nucleonic matter, we obtain stable hybrid star configurations 
for all the considered values of the gluon condensate and for the two considered sets 
of hyperon coupling constants.   
Notice that the hybrid star branch shrinks as $G_{2}$ is increased. 
Additionally, in the case $x_{\sigma}=0.6$, hybrid stars are possible only for  
$G_2  \lesssim 0.013 \, {\rm GeV}^4$.  In fact, for  $G_2  \gtrsim 0.013 \, {\rm GeV}^4$ 
the baryon number density $\rho_1$ for the onset of the mixed phase is larger than 
the central baryon number density $\rho_c^{HS}$ of the maximum mass pure hadronic star 
(see also Fig. \ref{rho_g2_xs06}).   

The properties of the maximum mass configuration for hybrid star sequences varying  
$G_{2}$ and the hyperon coupling constants are summarized in Table \ref{t:mass} 
together with those for pure nucleonic and hyperonic star sequences. 

We next plot in Fig. \ref{rho_g2_xs08} (case with $x_{\sigma} = 0.8$)  and 
in Fig. \ref{rho_g2_xs06}  (case with $x_{\sigma} = 0.6$) the quark-hadron phase transition boundaries in $\beta$-stable matter as a function of  $G_2$.  
As before (Fig. \ref{rho_g2_N}) we denote with $\rho_1$ ($\rho_2$)  
the density for the onset of the quark-hadron mixed phase (pure quark phase).  
The curve labeled  $\rho_c^{Hyb}$ represents the central density of the 
maximum mass hybrid star. 
The horizontal dashed line represents the value of the central density $\rho_c^{HS}$ 
of the maximum mass pure hadronic star (i.e. hyperonic star).   

As one can see, comparing the results in Figs. \ref{rho_g2_xs08} and \ref{rho_g2_xs06}  
with those reported in Fig. \ref{rho_g2_N},  for ''low'' values of the gluon condensate 
(i.e. $G_2  \lesssim 0.008 \, {\rm GeV}^4$), the densities $\rho_1$ and  $\rho_2$ 
are unaffected by the presence of hyperons.   
When $G_2  \gtrsim 0.008 \, {\rm GeV}^4$ the inclusion of hyperons produces a 
sizeable increase of the density $\rho_1$ and reduces the extension of the range 
$\rho_2 \, \textendash \,  \rho_1$ of the mixed phase.   
In particular, in the case $x_{\sigma} = 0.6$ (Fig. \ref{rho_g2_xs06}), as we have already 
pointed out, $\rho_1 > \rho_c^{HS}$ when  $G_2  \gtrsim 0.013 \, {\rm GeV}^4$, 
thus no deconfinement phase transition occurs in pure hyperonic stars.  

The hybrid star maximum mass $M_{max}$ (continuous line) and the 
mass $M_1 = M(\rho_1)$ (dashed line) as a function of  $G_2$ are plotted 
in Fig. \ref{Mmax_g2_xs08} (case with  $x_{\sigma} = 0.8$)  and 
in Fig. \ref{Mmax_g2_xs06} (case with  $x_{\sigma} = 0.6$).   
From the lower bound of the measured mass of PSR~J1614-2230, 
in the case with  $x_{\sigma} = 0.8$, we obtain $G_{2} \gtrsim 0.0127~{\rm GeV}^4$.  
Notice that in the case with  $x_{\sigma} = 0.6$,  
hybrid stars ($G_2  \lesssim 0.013 \, {\rm GeV}^4$) or pure hyperonic stars 
($G_2  \gtrsim 0.013 \, {\rm GeV}^4$) are not compatible  
the lower bound of the measured mass of PSR~J1614-2230.   

\begin{figure}
\centering
\includegraphics[scale=0.35]{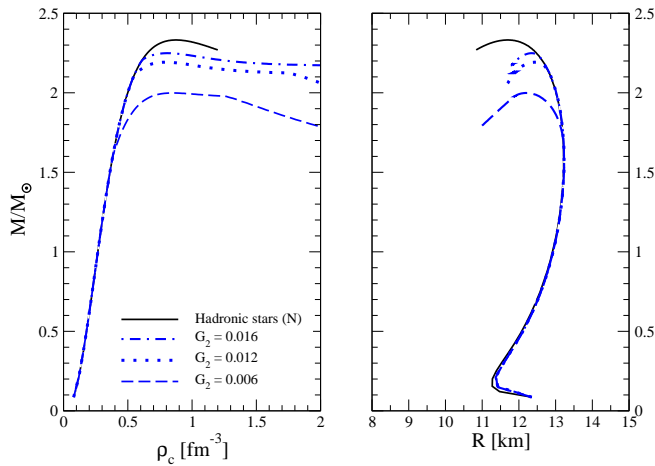}
\caption{Same as Fig. \ref{Mrho+MR_N}  but with  $V_1 = 0.10 \, {\rm GeV}$.} 
\label{Mrho+MR_N_V1=0.1}
\end{figure}

To explore the influence of the large-distance static $Q\bar{Q}$ potential $V_1$ 
on the stellar properties, we report in Fig. \ref{Mrho+MR_N_V1=0.1} 
the stellar mass $M$ versus $\rho_{c}$ (left panel) and $M$ versus $R$ (right panel) 
in the case of pure nucleonic stars (continuous line) and of hybrid stars 
for different $G_{2}$ values and  $V_1 = 0.10 \, {\rm GeV}$.  
Once again we get stable hybrid star configurations for all the considered 
values $G_2$, with maximum masses ranging from  
$M_{max} = 2.00 \, M_\odot$ (case with $G_{2} = 0.006 \, {\rm GeV}^4$) to  
$M_{max} = 2.25 \, M_\odot$ ($G_{2} = 0.0016 \, {\rm GeV}^4$).  
Comparing the results in Fig. \ref{Mrho+MR_N_V1=0.1} with those reported  
in Fig. \ref{Mrho+MR_N} (case with $V_1 = 0.01 \, {\rm GeV}$), 
we clearly see that a larger value of $V_1$ reduces the extent the hybrid star branch, 
shifts it to larger densities (see also results in Fig. \ref{fig2}) and produces hybrid stars 
with a larger maximum mass. 
Notice that, in this case the calculated $M_{max}$ is compatible with  
the lower bound of the measured mass of PSR~J1614-2230 for all the considered 
values of the gluon condensate (i.e. $G_2  \gtrsim 0.006 \, {\rm GeV}^4$).\\  

We also considered stellar models with $V_1 = 0.50 \, {\rm GeV}$ and 
$V_1 = 0.85 \, {\rm GeV}$. For these values of $V_1$ no phase transition occurs in neutron 
stars (i.e. $\rho_1 > \rho_c^{HS}$), thus in this case PSR~J1614-2230 would be 
a pure nucleonic star. 

\begin{figure}
\centering
\includegraphics[scale=0.35]{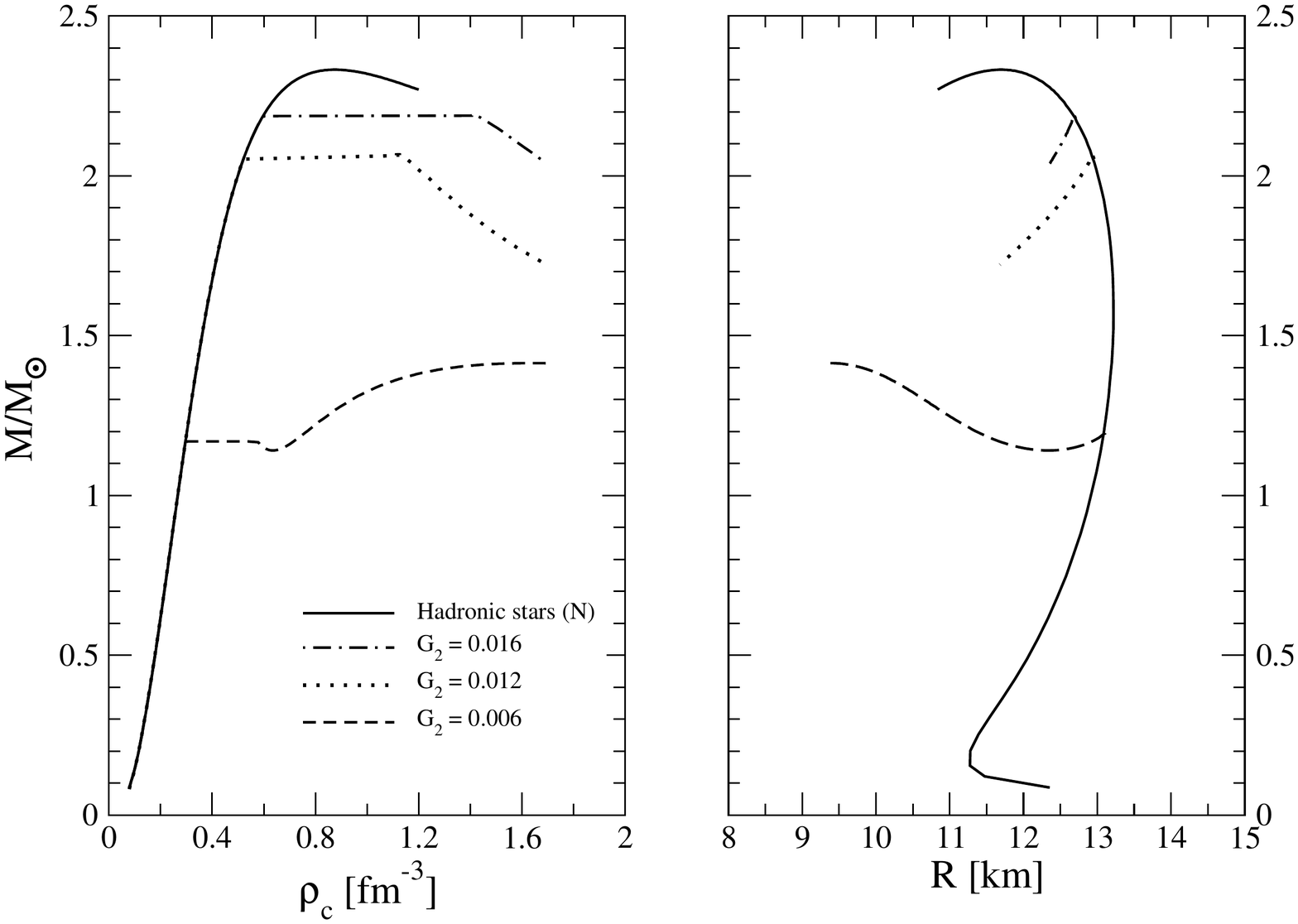}
\caption{Same as Fig. \ref{Mrho+MR_N} but  using the Maxwell construction to model 
the hadron-quark phase transition.}
\label{Mrho+MR_N_maxwell}
\end{figure} 

Finally to explore the role of the phase transition treatment, 
we plot in Fig. \ref{Mrho+MR_N_maxwell}, making use of the Maxwell construction, 
the stellar gravitational mass $M$ versus $\rho_{c}$ (left panel) and the 
mass versus radius $R$ (right panel) in the case of pure nucleonic stars (continuous line)  
and of hybrid stars for $V_{1} = 0.01 \, {\rm GeV}$ and for three different $G_{2}$ values.  
Stable hybrid stars exist in the case of $G_{2} = 0.006 \, {\rm GeV}^4$ 
having a maximum mass configuration with 
$M_{max} = 1.41 \, M_\odot$, $\rho_c^{Hyb} = 1.66 \, {\rm fm}^{-3}$, 
and $R = 9.35\, {\rm km}$.   
For the other two considered values of the gluon condensate  
($G_{2} = 0.012 \, {\rm GeV}^4$ and  $0.016 \, {\rm GeV}^4$) 
no stable hybrid star can be formed. In fact, an instability develops as soon as the 
stellar central density equals the critical density for the quark deconfinement transition.   
These unstable hybrid stars are those represented in Fig. \ref{Mrho+MR_N_maxwell}  
by the decreasing branch of the $M(\rho_c)$ curve  and by the configurations 
on the left of the cusp in the mass-radius curve.     
The results in Fig. \ref{Mrho+MR_N_maxwell} are in agreement with those reported  
in Ref. \cite{Baldo} where, making use of the Maxwell construction, stable hybrid star 
configurations were found only for $0.006 \, {\rm GeV}^4 < G_{2} < 0.007 \, {\rm GeV}^4$. 
Thus comparing the results in Fig. \ref{Mrho+MR_N_maxwell} with the analogous results 
in Fig. \ref{Mrho+MR_N}, but obtained using  the Gibbs construction, 
we deduce that the stability of hybrid star equlilibrium configurations within the field correlator method 
is related to the modeling of the deconfinement phase transition rather than to the confining features 
of the quark matter model \cite{Baldo}.     
We have verified that this conclusion is valid also in the case the hadronic phase contains  hyperons (NY matter with  $x_{\sigma} = 0.6$ and 0.8).

\section{Lattice QCD calculations and measured neutron star masses}
\label{Tc}   
  
\begin{figure}
\centering
\includegraphics[scale=0.33]{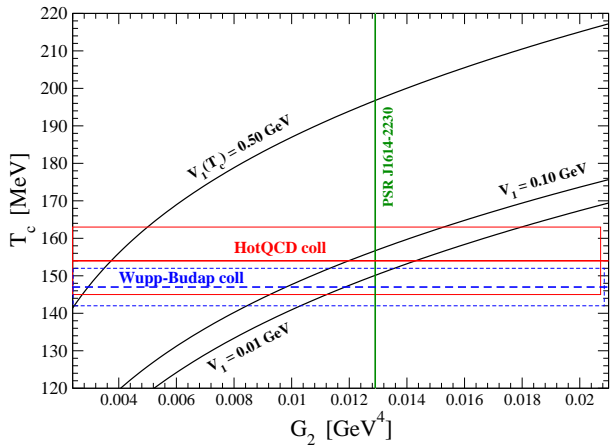}
\caption{(color online) Deconfinement transition temperature $T_c$ at $\mu_b = 0$.  
The curve labeled with $V_1(T_c) = 0.5 \,{\rm GeV}$ reproduces the FCM results of 
Ref.~\cite{ST07} for a fixed value $V_1(T_c) = 0.5 \,{\rm GeV}$ of the large-distance 
static $Q\bar{Q}$ potential.  
The curve labeled with $V_1= 0.01\,{\rm GeV}$ ($V_1= 0.10\,{\rm GeV}$)  
corresponds to the transition temperature at $\mu_b = 0$ obtained   
solving numerically Eqs.~(\ref{Tc0}) and  (\ref{v1_T}) for the case 
$V_1(0) = 0.01\,{\rm GeV}$ [$V_1(0) = 0.10\,{\rm GeV}$].     
The horizontal heavy and thin lines represent respectively the central 
value and the error estimate of lattice QCD calculations. 
In particular, the (red) continuous lines refer to the calculations \cite{baz12} 
of the HotQCD Collaboration $T_c = (154 \pm 9)\, {\rm MeV}$;  
the (blue) short-dashed lines refer to the calculations~\cite{bors10}  
of the Wuppertal-Budapest Collaboration  $T_c =  (147 \pm 5)\, {\rm MeV}$. 
Finally, the vertical green line represents the lower limit for $G_2$ which is compatible 
with the lower bound of the measured mass of PSR~J1614-2230 for the case 
$V_1(0) = 0.01\,{\rm GeV}$.}   
\label{Tc_G2}
\end{figure}

Within the FCM the deconfinement transition temperature $T_c$ at $\mu_b = 0$ 
reads \cite{ST07} 
\be
\label{Tc0}
T_c = \frac{a_0}{2} G_2^{1/4} \Bigg( 1 + \sqrt{
1 + \frac{V_1(T_c)}{2a_0 G_2^{1/4}}} \,~  \Bigg)\, , 
\ee
with $a_0 = (3 \pi^2/768)^{1/4}$ in the case of three flavors.\\  
In their analysis the authors of Ref.~\cite{ST07} assume  $V_1(T_c) = 0.5 \,{\rm GeV}$, 
thus $T_c$ in Eq. (\ref{Tc0}) is a simple function of $G_2$ and is represented 
in Fig.~\ref{Tc_G2} by the curve labeled $V_1(T_c) = 0.5 \,{\rm GeV}$.   
This result can hence be compared with lattice QCD calculations of $T_c$ giving the possibility 
to extract the range of values for the gluon condensate compatible with lattice results. 
This comparison has been done by the authors of Ref.~\cite{ST07}, and it is done in the 
present work in Fig.~\ref{Tc_G2}, where we consider the recent lattice QCD calculations carried out 
by the HotQCD Collaboration \cite{baz12} giving $T_c = (154 \pm 9)\, {\rm MeV}$ (red continuous lines)  
and by the Wuppertal-Budapest Collaboration \cite{bors10} giving $T_c =  (147 \pm 5)\, {\rm MeV}$ 
(blue short-dashed lines). 
As one can see, the comparison with lattice QCD calculations of $T_c$ restricts 
the gluon condensate in a rather narrow range  $G_2 = 0.0025$--$0.0050 \, \rm{GeV}^4$.  

Next to verify if these values of $G_2$ are compatible with those extracted in 
Sec.~\ref{NSstr} from hybrid star calculations and measured neutron star masses, 
we need to relate the parameter $V_1 \equiv V_1(0)$ entering in the zero temperature 
EOS of the quark phase with  $V_1(T_c)$ in Eq.(\ref{Tc0}). 
To this end, one can integrate Eq.(\ref{v1}) using the nonperturbative  
contribution (\ref{d1nonpt}) to the color electric correlator $D_1^E(x)$ and assuming 
that the normalization factor $D_1^E(0)$ does not depend on temperature. 
The latter assumption is supported, up to temperatures very near to $T_c$,  
by lattice calculations \cite{pisa1,pisa2,pisa3}.   
Therefore  one gets  
\be
\label{v1_T}
V_1(T)  = V_1(0) 
\bigg\{1-\frac{3}{2} \frac{\lambda T}{\hbar c} + 
\frac{1}{2} \bigg(1 + 3 \frac{\lambda T}{\hbar c}\bigg) e^{- \frac{\hbar c}{\lambda T}} 
\bigg\}  \, . 
\ee
Thus $V_1(T_c) = 0.5 \,{\rm GeV}$ corresponds to  $V_1(0) = 0.85 \,{\rm GeV}$ 
to be used in the $T = 0$ EOS of the quark phase. 
In this case, as we found in Sec.~\ref{NSstr}, no phase transition occurs in neutron 
stars (i.e $\rho_1 > \rho_c^{NS}$) for all the considered values of $G_2$. 
Thus for these values of the EOS parameters PSR~J1614-2230 would be a pure nucleonic star. 
  
We can also evaluate the FCM transition temperature at $\mu_b=0$ corresponding 
to the case  $V_1(0) = 0.01 \,{\rm GeV}$ used in Sec.~\ref{NSstr} for hybrid star 
calculations with the $T=0$ FCM equation of state.     
To this purpose we solve numerically Eqs.~(\ref{Tc0}) and (\ref{v1_T}) and 
we obtain the results represented in Fig.~\ref{Tc_G2} by the curve labeled 
$V_1 = 0.01 \,{\rm GeV}$. The comparison of these results with 
lattice QCD calculations \cite{bors10,baz12} of $T_c$ restricts the gluon 
condensate in the range  $G_2 = 0.0103$--$0.0180 \, \rm{GeV}^4$.  
Coming now to the astrophysical constraints on the gluon condensate,  
the vertical green line in Fig.~\ref{Tc_G2}  represents the lower limit for $G_2$ 
which is compatible, in the case $V_1(0) = 0.01 \,{\rm GeV}$, with the lower bound 
of the measured mass of PSR~J1614-2230 (see Sec.~\ref{NSstr}).\\ 
A similar analysis can be done for the case $V_1(0) = 0.10 \,{\rm GeV}$. 
Now the comparison between the FCM transition temperature at $\mu_b=0$ (curve 
labeled $V_1 = 0.10 \,{\rm GeV}$ in Fig.~\ref{Tc_G2}) and lattice QCD 
calculations of the same quantity gives 
$G_2 = 0.0085$--$0.0153 \, \rm{GeV}^4$, 
whereas one gets $G_2  \geq 0.006 \, {\rm GeV}^4$ from the lower bound of 
the measured mass of PSR~J1614-2230.\\  
 
We thus find that the values of the gluon condensate extracted within the FCM from 
lattice QCD calculations of the deconfinement transition temperature at zero baryon 
chemical potential are compatible with the value of the same quantity extracted 
from measured neutron star masses.

\section{Summary and Conclusions}
\label{concl}
 
 In this article we have studied the hadron-quark deconfinement phase transition 
in $\beta$-stable matter and the structural properties of hybrid stars using an EOS 
for the quark phase derived from the field correlator method extended to finite 
baryon chemical potential. This EOS model was parametrized in terms of the gluon 
condensate $G_2$ and of the large-distance static $Q\bar{Q}$ potential $V_1$ at zero temperature.  
For the hadronic phase we utilized the GM1 parametrization of the nonlinear relativistic 
mean field model, and we have considered pure nucleonic matter as well as 
hyperonic matter with a large hyperon fraction ($x_\sigma$=0.6),  
and a small hyperon fraction ($x_\sigma=0.8$).  

 We found that increasing the value of the gluon condensate $G_{2}$ caused 
a shift of the phase transition to larger baryon densities.  
Moreover, for the case $V_1 = 0.01 \,{\rm GeV}$,  
when $G_2  \gtrsim 0.008 \, {\rm GeV}^4$ the inclusion of hyperons 
produced a considerable increase of the mixed phase onset density $\rho_1$ and 
reduced the extension of the range  $\rho_2\,\textendash \,\rho_1$ of the mixed phase.    
In particular, in the case $x_{\sigma} = 0.6$, when $G_2  \gtrsim 0.013 \, {\rm GeV}^4$ 
no deconfinement phase transition occurred in pure hyperonic stars.  

Applying the Gibbs construction to model the phase transition, we obtained stable 
hybrid star configurations for all the values of the gluon condensate fulfilling the 
condition $\rho_1(G_2) < \rho_c^{HS}(G_2)$ (i.e. the deconfinement transition 
can occur in pure hadronic stars). 
We found that the hybrid star branch shrank as $G_{2}$ was increased.  
 
We have established that the values of the gluon condensate extracted within the FCM 
from lattice QCD calculations of the deconfinement transition temperature at $\mu_b = 0$ 
were consistent with the value of the same quantity derived by the mass measurement 
of PSR~J1614-2230. 
The FCM thus provides a powerful tool to link numerical calculations of QCD on a space-time lattice 
with neutron stars physics.   


\end{document}